\documentclass[a4paper, 11pt]{article}
\pdfoutput=1

\usepackage{jheppub}
\usepackage{afterpage}
\usepackage{ifthen}

\newenvironment{pagefigure}{\begin{figure}[!p]}{\afterpage\clearpage\end{figure}}

\newcommand{\Dmq}{\Delta m^2}
\newcommand{\Nuc}[2][]{{\ensuremath{\ifthenelse{\equal{#1}{}}{}{\mbox{}^{#1}}\text{#2}}}}
\newcommand{\eVq}{\ensuremath{\text{eV}^2}}
\renewcommand{\Re}{\mathop{\rm Re}}

\title{Global fit to three neutrino mixing: critical look
at present precision}

\author[a,b]{M.~C.~Gonzalez-Garcia,}
\affiliation[a]{C.N.~Yang Institute for Theoretical Physics,
  State University of New York at Stony Brook,
  Stony Brook, NY 11794-3840, USA}
\affiliation[b]{Instituci\'o Catalana de Recerca i Estudis Avan\c{c}ats (ICREA),
  Departament d'Estructura i Constituents de la Mat\`eria and
  Institut de Ciencies del Cosmos,
  Universitat de Barcelona, Diagonal 647, E-08028 Barcelona, Spain}
\emailAdd{concha@insti.physics.sunysb.edu}

\author[c]{Michele Maltoni,}
\affiliation[c]{Instituto de F\'{\i}sica Te\'orica UAM/CSIC, Calle de Nicol\'as
  Cabrera 13--15, Universidad Aut\'onoma de Madrid, Cantoblanco,
  E-28049 Madrid, Spain}
\emailAdd{michele.maltoni@csic.es}

\author[d]{Jordi Salvado,}
\affiliation[d]{Departament d'Estructura i Constituents de la Mat\`eria and
  Institut de Ciencies del Cosmos,
  Universitat de Barcelona, 647 Diagonal, E-08028 Barcelona, Spain}
\emailAdd{jsalvado@ecm.ub.es}

\author[e]{Thomas Schwetz}
\affiliation[e]{Max-Planck-Institut f\"ur Kernphysik, Saupfercheckweg 1,
  69117 Heidelberg, Germany}
\emailAdd{schwetz@mpi-hd.mpg.de}

\abstract{We present an up-to-date global analysis of solar,
  atmospheric, reactor, and accelerator neutrino data in the framework
  of three-neutrino oscillations. We provide results on the
  determination of $\theta_{13}$ from global data and discuss the
  dependence on the choice of reactor fluxes. We study in detail the
  statistical significance of a possible deviation of $\theta_{23}$
  from maximal mixing, the determination of its octant, the ordering
  of the mass states, and the sensitivity to the CP violating phase,
  and discuss the role of various complementary data sets in those
  respects.}

\preprint{IFT-UAM/CSIC-12-88, YITP-SB-12-33}


\begin{document}

\maketitle

\section{Introduction}

It is now an established fact that neutrinos are massive and leptonic
flavors are not symmetries of Nature~\cite{Pontecorvo:1967fh,
  Gribov:1968kq}.  In the last decade this picture has become fully
proved thanks to the upcoming of a set of precise experiments. In
particular, the results obtained with solar and atmospheric neutrinos
have been confirmed in experiments using terrestrial beams: neutrinos
produced in nuclear reactors and accelerators facilities have been
detected at distances of the order of hundreds of
kilometers~\cite{GonzalezGarcia:2007ib}.
The minimum joint description of all these data requires mixing among
all the three known neutrinos ($\nu_e$, $\nu_\mu$, $\nu_\tau$), which
can be expressed as quantum superpositions of three massive states
$\nu_i$ ($i=1,2,3$) with masses $m_i$.  This implies the presence of a
leptonic mixing matrix in the weak charged current
interactions~\cite{Maki:1962mu, Kobayashi:1973fv} which can be
parametrized as:
\begin{equation}
  \label{eq:matrix}
  U =
  \begin{pmatrix}
    c_{12} c_{13}
    & s_{12} c_{13}
    & s_{13} e^{-i\delta_\text{CP}}
    \\
    - s_{12} c_{23} - c_{12} s_{13} s_{23} e^{i\delta_\text{CP}}
    & \hphantom{+} c_{12} c_{23} - s_{12} s_{13} s_{23}
    e^{i\delta_\text{CP}}
    & c_{13} s_{23} \hspace*{5.5mm}
    \\
    \hphantom{+} s_{12} s_{23} - c_{12} s_{13} c_{23} e^{i\delta_\text{CP}}
    & - c_{12} s_{23} - s_{12} s_{13} c_{23} e^{i\delta_\text{CP}}
    & c_{13} c_{23} \hspace*{5.5mm}
  \end{pmatrix},
\end{equation}
where $c_{ij} \equiv \cos\theta_{ij}$ and $s_{ij} \equiv
\sin\theta_{ij}$.  In addition to the Dirac-type phase
$\delta_\text{CP}$, analogous to that of the quark sector, there are
two physical phases associated to the Majorana character of neutrinos,
which however are not relevant for neutrino
oscillations~\cite{Bilenky:1980cx, Langacker:1986jv} and are therefore
omitted in the present work.  Given the observed hierarchy between the
solar and atmospheric mass-squared splittings there are two possible
non-equivalent orderings for the mass eigenvalues, which are
conventionally chosen as
\begin{align}
  \label{eq:normal}
  \Dmq_{21} &\ll \hphantom{+} (\Dmq_{32} \simeq \Dmq_{31} > 0) \,;
  \\
  \label{eq:inverted}
  \Dmq_{21} &\ll -(\Dmq_{31} \simeq \Dmq_{32} < 0) \,,
\end{align}
with $\Dmq_{ij} \equiv m_i^2 - m_j^2$.  As it is customary we refer to
the first option, Eq.~\eqref{eq:normal}, as Normal ordering (NO), and
to the second one, Eq.~\eqref{eq:inverted}, as Inverted ordering (IO);
in this form they correspond to the two possible choices of the sign
of $\Dmq_{31}$.\footnote{In the following we adopt the (arbitrary)
  convention of reporting results for $\Dmq_{31}$ for NO and
  $\Dmq_{32}$ for IO, \textit{i.e.}, we always use the one which has
  the larger absolute value.} In this convention the angles
$\theta_{ij}$ can be taken without loss of generality to lie in the
first quadrant, $\theta_{ij} \in [0, \pi/2]$, and the CP phase
$\delta_\text{CP} \in [0, 2\pi]$.

Within this context, $\Dmq_{21}$, $|\Dmq_{31}|$, $\theta_{12}$, and
$\theta_{23}$ are relatively well determined, whereas till this year
only an upper bound was derived for the mixing angle $\theta_{13}$ and
barely nothing was known on the CP phase $\delta_\text{CP}$ and on the
sign of $\Dmq_{31}$. This situation has dramatically changed with the
data from the reactor experiments Daya Bay~\cite{An:2012eh},
Reno~\cite{Ahn:2012nd}, and Double Chooz~\cite{Abe:2011fz}, which
together with the increased statistics of long-baseline experiments
T2K~\cite{Abe:2011sj} and MINOS~\cite{Adamson:2011qu} have provided a
a clear determination of the last unknown mixing angle $\theta_{13}$.
With these results at hand a first-order picture of the three-flavour
lepton mixing matrix has emerged. Its precise determination, as well
as that of the mass differences, can only be made by statistically
combining the results of the oscillation searches.

In this article, we present an up-to-date global analysis of solar,
atmospheric, reactor and accelerator neutrino data in the framework of
three-neutrino oscillations with the following outline.
Sec.~\ref{sec:global} contains the results of the global analysis and
the extracted ranges of the oscillation parameters and the
corresponding mixing matrix. In Sec.~\ref{sec:t13} we study the
present determination of $\theta_{13}$ and in particular we discuss
the uncertainty associated to the choice of reactor fluxes.
Sec.~\ref{sec:t23del} focuses on our knowledge of $\theta_{23}$ and
$\delta_\text{CP}$ with special emphasis on the present limitations of
the statistical significance of the possible deviation of
$\theta_{23}$ from maximal, the determination of its octant and the
sensitivity to the CP violating phase. In order to address these
issues we study the role played by a potential complementarity of long
baseline accelerator and reactor experiments
(Sec.~\ref{sec:beam-react}) and by the atmospheric neutrino results
(Sec.~\ref{sec:atm}). Finally, in Sec.~\ref{sec:dis} we summarize and
discuss our results. Future updates of this analysis will be provided at the
website Ref.~\cite{nufit}. Alternative recent global fits have been
presented in Refs.~\cite{Fogli:2012ua, Tortola:2012te}, see
also~\cite{Machado:2011ar, Bergstrom:2012yi}.

\section{Oscillation Parameters: Results of the Global Analysis}
\label{sec:global}

We include in our global analysis the results from Super-Kamiokande
atmospheric neutrino data from phases SK1--4~\cite{skatm:nu2012}
(with addition of the 1097 days of phase SK4 over their published
results on phases SK1--3~\cite{Wendell:2010md}).  For what concerns
long baseline accelerator experiments (LBL), we combine the energy
distribution of $\nu_\mu$ disappearance events from
K2K~\cite{Ahn:2006zza} with that obtained by MINOS in both $\nu_\mu$
($\bar\nu_\nu$) disappearance and $\nu_e$ ($\bar\nu_e$) appearance
with $10.8~(3.36) \times 10^{20}$ protons on target
(pot)~\cite{minos:nu2012} (which update their published
results~\cite{Adamson:2012rm, Adamson:2011qu}), and T2K $\nu_e$
appearance and $\nu_\mu$ disappearance data with phases 1-3,
$3.01\times 10^{20}$ pot~\cite{t2k:ichep2012} (a factor $\sim$ 2
increase with respect to their published results~\cite{Abe:2011sj}),
and phases 1-2, $1.43\times 10^{20}$ pot~\cite{Abe:2012gx,
  t2k:nu2012}, respectively.

For oscillation signals at reactor experiments we include data from
the finalized experiments CHOOZ~\cite{Apollonio:1999ae} (energy
spectrum data) and Palo Verde~\cite{Piepke:2002ju} (total rate)
together with the recent spectrum from Double Chooz with 227.9 days
live time~\cite{Abe:2012uy, dc:nu2012}, and the total even rates in
the near and far detectors in Daya Bay~\cite{dayabay:nu2012} with 126
live days of data (a factor 3 increase over their published
results~\cite{An:2012eh}) and Reno with 229 days of
data-taking~\cite{Ahn:2012nd}.  We also include the observed energy
spectrum in KamLAND data sets DS-1 and DS-2~\cite{Gando:2010aa} with a
total exposure of $3.49\times 10^{32}$ target-proton-year (2135 days).

Finally in the analysis of solar neutrino experiments we include the
total rates from the radiochemical experiments
Chlorine~\cite{Cleveland:1998nv}, Gallex/GNO~\cite{Kaether:2010ag} and
SAGE~\cite{Abdurashitov:2009tn}. For real-time experiments we include
the 44 data points of the electron scattering (ES) Super-Kamiokande
phase I (SK1) energy-zenith spectrum~\cite{Hosaka:2005um} and the data
from the three phases of SNO~\cite{Aharmim:2007nv, Aharmim:2005gt,
  Aharmim:2008kc}, including the results on the low energy threshold
analysis of the combined SNO phases I--III~\cite{Aharmim:2011vm}.  We
also include the main set of the 740.7 days of Borexino
data~\cite{Bellini:2011rx} as well as their high-energy spectrum from
246 live days~\cite{Collaboration:2008mr}.

\begin{pagefigure}\centering
  \includegraphics[width=0.81\textwidth]{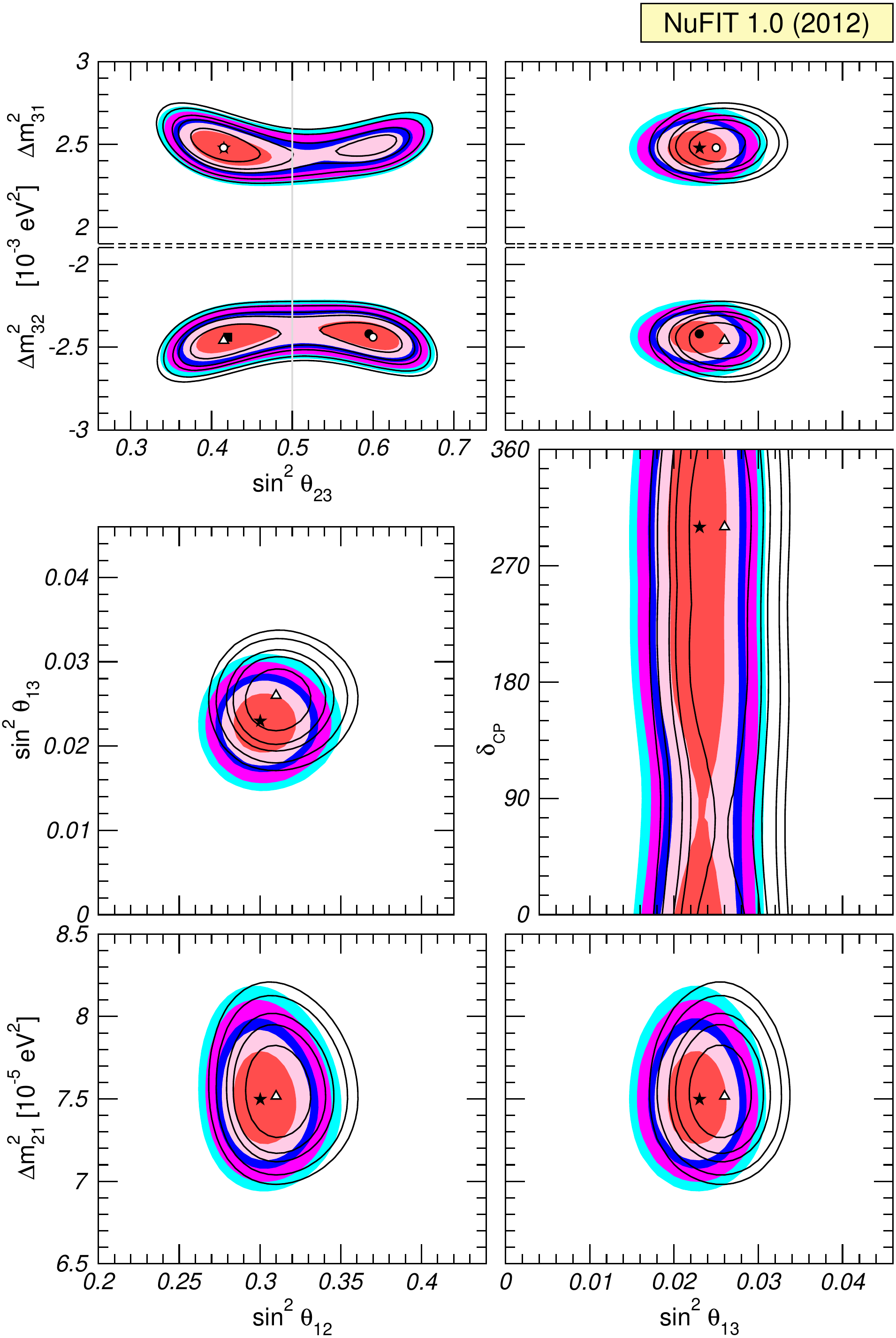}
  \caption{Global $3\nu$ oscillation analysis.  Each panels shows
    two-dimensional projection of the allowed six-dimensional region
    after marginalization with respect to the undisplayed parameters.
    The different contours correspond to the two-dimensional allowed
    regions at $1\sigma$, 90\%, $2\sigma$, 99\% and $3\sigma$ CL
    (2~dof).  Results for different assumptions concerning the
    analysis of data from reactor experiments are shown: full regions
    correspond to analysis with the normalization of reactor fluxes
    left free and data from short-baseline (less than 100 m) reactor
    experiments are included.  For void regions short-baseline reactor
    data are not included but reactor fluxes as predicted
    in~\cite{Huber:2011wv} are assumed. Note that as atmospheric
    mass-squared splitting we use $\Dmq_{31}$ for NO and $\Dmq_{32}$
    for IO.}
  \label{fig:region-glob}
\end{pagefigure}

\begin{pagefigure}\centering
  \includegraphics[width=0.86\textwidth]{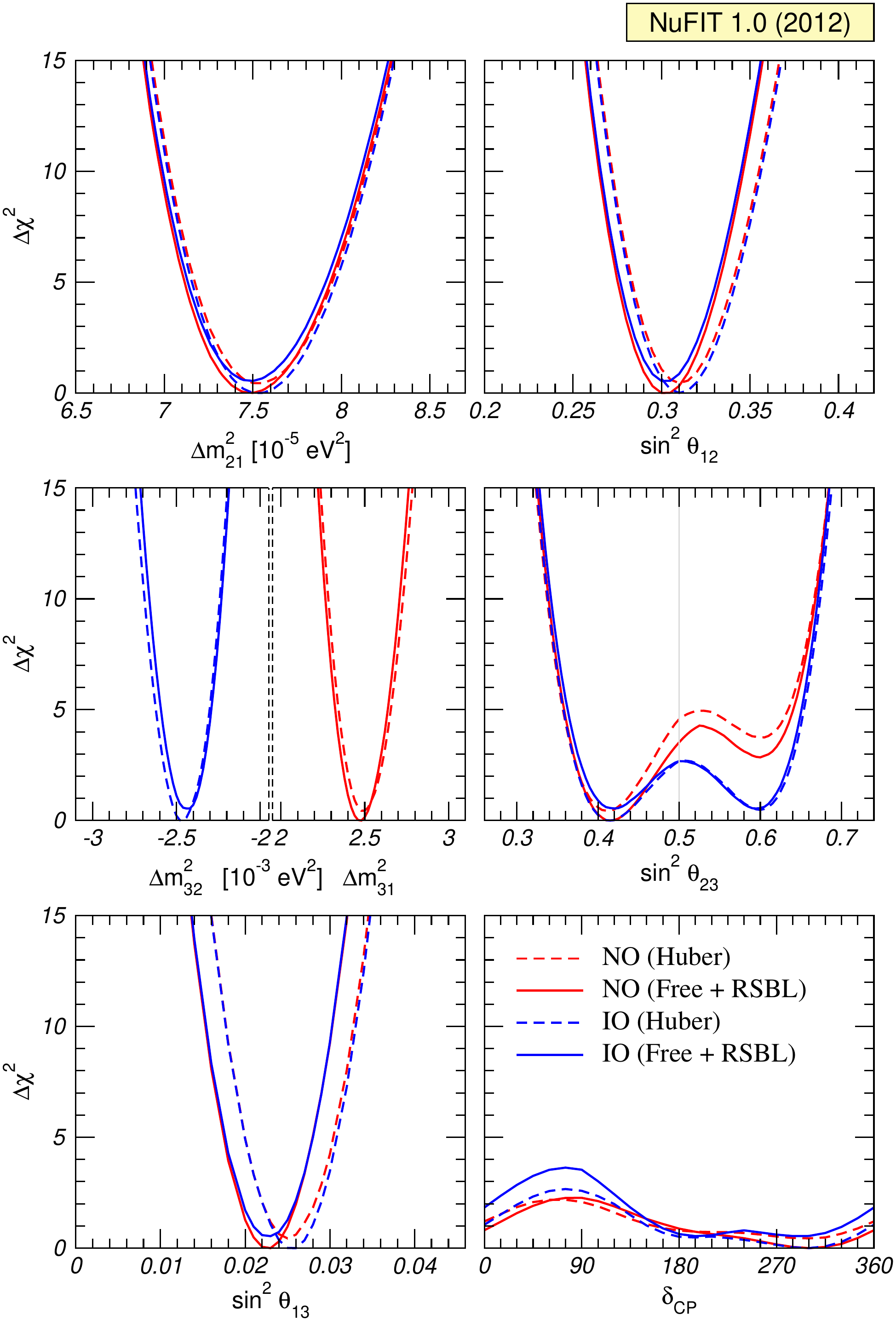}
  \caption{Global $3\nu$ oscillation analysis.  The red (blue) curves
    are for Normal (Inverted) Ordering.  Results for different
    assumptions concerning the analysis of data from reactor
    experiments are shown: for solid curves the normalization of
    reactor fluxes is left free and data from short-baseline (less
    than 100 m) reactor experiments are included.  For dashed curves
    short-baseline data are not included but reactor fluxes as
    predicted in~\cite{Huber:2011wv} are assumed. Note that as
    atmospheric mass-squared splitting we use $\Dmq_{31}$ for NO and
    $\Dmq_{32}$ for IO.}
  \label{fig:chisq-glob}
\end{pagefigure}

\begin{table}\centering
  \footnotesize
  \begin{tabular}{@{}l|cc|cc@{}}
    \hline\hline
    & \multicolumn{2}{c|}{Free Fluxes + RSBL}
    & \multicolumn{2}{c@{}}{Huber Fluxes, no RSBL}
    \\
    \hline
    & bfp $\pm 1\sigma$ & $3\sigma$ range
    & bfp $\pm 1\sigma$ & $3\sigma$ range
    \\
    \hline
    \rule{0pt}{4mm}
    $\sin^2\theta_{12}$
    & $0.302_{-0.012}^{+0.013}$ & $0.267 \to 0.344$
    & $0.311_{-0.013}^{+0.013}$ & $0.273 \to 0.354$
    \\[1mm]
    $\theta_{12}/^\circ$
    &$33.36_{-0.78}^{+0.81}$ & $31.09 \to 35.89$
    &$33.87_{-0.80}^{+0.82}$ & $31.52 \to 36.49$
    \\[3mm]
    $\sin^2\theta_{23}$
    & $0.413_{-0.025}^{+0.037} \oplus 0.594_{-0.022}^{+0.021}$ & $0.342 \to 0.667$
    & $0.416_{-0.029}^{+0.036} \oplus 0.600_{-0.026}^{+0.019}$ & $0.341 \to 0.670$
    \\[1mm]
    $\theta_{23}/^\circ$
    & $40.0_{-1.5}^{+2.1} \oplus 50.4_{-1.3}^{+1.3}$ & $35.8 \to 54.8$
    & $40.1_{-1.6}^{+2.1} \oplus 50.7_{-1.5}^{+1.2}$ & $35.7 \to 55.0$
    \\[3mm]
    $\sin^2\theta_{13}$
    & $0.0227_{-0.0024}^{+0.0023}$ & $0.0156 \to 0.0299$
    & $0.0255_{-0.0024}^{+0.0024}$ & $0.0181 \to 0.0327$
    \\
    $\theta_{13}/^\circ$
    & $8.66_{-0.46}^{+0.44}$ & $7.19 \to 9.96$
    & $9.20_{-0.45}^{+0.41}$ & $\hphantom{0}7.73 \to 10.42$
    \\[3mm]
    $\delta_\text{CP}/^\circ$
    & $300_{-138}^{+66}$  & $\hphantom{00}0 \to 360$
    & $298_{-145}^{+59}$  & $\hphantom{00}0 \to 360$
    \\[3mm]
    $\dfrac{\Dmq_{21}}{10^{-5}~\eVq}$
    & $7.50_{-0.19}^{+0.18}$ & $7.00 \to 8.09$
    & $7.51_{-0.15}^{+0.21}$ & $7.04 \to 8.12$
    \\[3mm]
     $\dfrac{\Dmq_{31}}{10^{-3}~\eVq}$ (N)
    & $+2.473_{-0.067}^{+0.070}$ & $+2.276 \to +2.695$
    & $+2.489_{-0.051}^{+0.055}$ & $+2.294 \to +2.715$
     \\[3mm]
     $\dfrac{\Dmq_{32}}{10^{-3}~\eVq}$ (I)
    & $-2.427_{-0.065}^{+0.042}$ & $-2.649 \to -2.242$
    & $-2.468_{-0.065}^{+0.073}$ & $-2.678 \to -2.252$
    \\[3mm]
    \hline\hline
  \end{tabular}
  \caption{Three-flavour oscillation parameters from our fit to global
    data after the Neutrino~2012 conference. For ``Free Fluxes +
    RSBL'' reactor fluxes have been left free in the fit and short
    baseline reactor data (RSBL) with $L \lesssim 100$~m are included;
    for ``Huber Fluxes, no RSBL'' the flux prediction
    from~\cite{Huber:2011wv} are adopted and RSBL data are not used in
    the fit.}
  \label{tab:results}
\end{table}

The results of the global analysis are shown in
Figs.~\ref{fig:region-glob} and~\ref{fig:chisq-glob} where we show
different projections of the allowed six-dimensional parameter
space. The results are shown for two choices of the reactor fluxes as
we will describe in more detail in the next section. The best fit
values and the derived ranges for the six parameters at the $1\sigma$
($3\sigma$) level are given in Tab.~\ref{tab:results}. For each
parameter the ranges are obtained after marginalizing with respect to
the other parameters.  For $\sin^2\theta_{23}$ the $1\sigma$ ranges
are formed by two disconnected intervals in which the first one
contains the absolute minimum and the second-one the secondary local
minimum. Note that we marginalize also over the type of the neutrino
mass ordering and the two local minima in $\sin^2\theta_{23}$ may
correspond to different orderings. As visible in
  Fig.~\ref{fig:chisq-glob}, for ``Free Fluxes + RSBL'' the best fit
  with $\sin^2\theta_{23} < 0.5$ is for NO and the local minimum with
  $\sin^2\theta_{23} > 0.5$ is for IO, while for ``Huber fluxes, no
  RSBL'' both minima are for IO. However, as also visible from the
  figure, the best fit values of the minima and the allowed range of
  $\sin^2\theta_{23}$ are not very sensitivite to the mass ordering.
The $3\sigma$ range is connected.  For $\Dmq_{31}$ ($\Dmq_{32}$) the
allowed ranges are formed by two disconnected intervals which
correspond to the two possible mass orderings.
From these results we conclude that:
\begin{enumerate}
\item The present global analysis disfavours $\theta_{13}=0$ with a
  $\Delta\chi^2 \approx 100$. This is mostly driven by the new reactor
  data from Daya Bay, Reno and Double Chooz (see
  Fig.~\ref{fig:chisq-t13} in the next section).

\item An uncertainty on $\theta_{13}$ at the level of $1\sigma$
  remains due to a tension between predicted reactor neutrino fluxes
  and data from reactor experiments with baselines less than 100~m, as
  we will discuss in detail in Sec.~\ref{sec:t13}.

\item Non-maximal $\theta_{23}$ is favoured at the level of $\sim
  2\sigma$ ($\sim 1.5 \sigma$) for Normal (Inverted) ordering for
  either choice of the reactor fluxes. We elaborate more on this issue
  in Sec.~\ref{sec:t23del}.

\item The statistical significance of the preference of the fit for the
  first octant of $\theta_{23}$ is $\leq 1.5 \sigma$ ($\leq 0.9
  \sigma$) for Normal (Inverted) ordering for either choice of the
  reactor fluxes.

\item When the normalization of reactor fluxes is left free and data
  from short-baseline (less than 100 m) reactor experiments are
  included, the absolute best-fit occurs for Normal ordering but the
  statistical significance of the preference Normal versus Inverted is
  $\leq 0.7 \sigma$.

\item The best fit occurs for Inverted ordering when reactor
  short-baseline data are not included and reactor fluxes as predicted
  in~\cite{Huber:2011wv} are assumed but the statistical significance
  of the preference Inverted versus Normal is $\leq 0.75 \sigma$.

\item The statistical significance of the effects associated with
  $\delta_\text{CP}$ is $\leq 1.5 \sigma $ ($\leq 1.75 \sigma$) for
  Normal (Inverted) ordering (see Sec.~\ref{sec:t23del}).
\end{enumerate}

From the global $\chi^2$ and following the procedure outlined in
Ref.~\cite{GonzalezGarcia:2003qf} one can derive the following
$3\sigma$ CL ranges on the magnitude of the elements of the leptonic
mixing matrix
\begin{equation}
  \label{eq:umatrix}
  |U| = \begin{pmatrix}
    0.795 \to 0.846 &\qquad
    0.513 \to 0.585 &\qquad
    0.126 \to 0.178
    \\
    0.205 \to 0.543 &\qquad
    0.416 \to 0.730 &\qquad
    0.579 \to 0.808
    \\
    0.215 \to 0.548 &\qquad
    0.409 \to 0.725 &\qquad
    0.567 \to 0.800
  \end{pmatrix} .
\end{equation}
By construction the derived limits in Eq.~\eqref{eq:umatrix} are
obtained under the assumption of the matrix $U$ being unitary.  In
other words, the ranges in the different entries of the matrix are
correlated due to the fact that, in general, the result of a given
experiment restricts a combination of several entries of the matrix,
as well as to the constraints imposed by unitarity.  As a consequence
choosing a specific value for one element further restricts the range
of the others.

\section{Determination of $\theta_{13}$ and Flux Uncertainties}
\label{sec:t13}

At present, reactor experiments with $L \sim 1$~km provide us with the
most precise determination of $\theta_{13}$.  Up to very recently the
interpretation of neutrino oscillation searches at nuclear power
plants was based on the calculations of the reactor $\bar\nu_e$ flux
from Ref.~\cite{Schreckenbach:1985ep}. Indeed, the observed rates at
all reactor experiments performed so-far at distances $L \lesssim
1$~km are consistent with these fluxes, therefore setting limits on
$\bar\nu_e$ disappearance. Over the last two years the flux of
$\bar\nu_e$ emitted from nuclear power plants has been
re-evaluated~\cite{Mueller:2011nm, Huber:2011wv}, yielding roughly 3\%
higher neutrino fluxes than assumed previously. As discussed in
Ref.~\cite{Mention:2011rk} this might indicate an anomaly in reactor
experiments at $L \lesssim 1$~km, which according to the new fluxes
observe a slight deficit. For the Chooz and Palo Verde experiments at
$L \simeq 1$~km a non-zero $\theta_{13}$ could lead to $\bar\nu_e$
disappearance accounting for the reduction of the rate.  However,
$\Dmq_{13}$ and $\theta_{13}$ driven oscillations will have no effect
in short-baseline (SBL) experiments with $L \lesssim 100$~m.

Motivated by this situation we follow here the approach of
Ref.~\cite{Schwetz:2011qt} and study the dependence of the determined
value of $\theta_{13}$ on the assumptions about the reactor fluxes,
see also~\cite{Ciuffoli:2012yd}. In particular we consider the impact
of allowing for a free normalization of the reactor fluxes
$f_\text{flux}$ and/or of including in the analysis the results of the
reactor experiments with $L\lesssim 100$ m
Bugey4~\cite{Declais:1994ma}, ROVNO4~\cite{Kuvshinnikov:1990ry},
Bugey3~\cite{Declais:1994su}, Krasnoyarsk~\cite{Vidyakin:1987ue,
  Vidyakin:1994ut}, ILL~\cite{Kwon:1981ua},
G\"osgen~\cite{Zacek:1986cu}, SRP~\cite{Greenwood:1996pb}, and
ROVNO88~\cite{Afonin:1988gx}, to which we refer as reactor
short-baseline experiments (RSBL).

\begin{figure}\centering
  \includegraphics[width=0.8\textwidth]{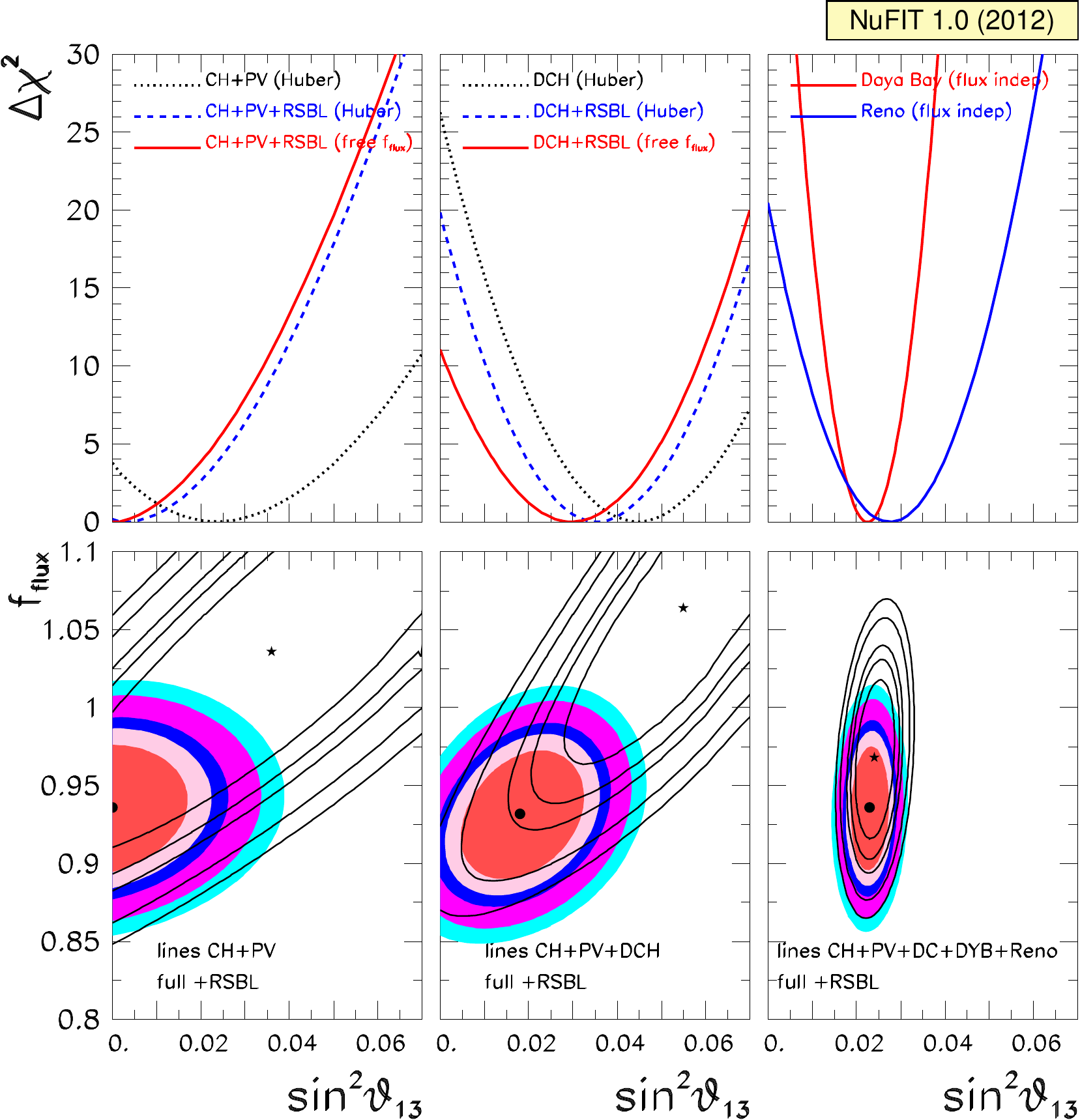}
  \caption {Upper: $\Delta\chi^2$ as a function of $\sin^2\theta_{13}$
    for the different reactor experiments and different assumptions on
    the fluxes as labeled in the figure.  In this figure we fix
    $\Dmq_{31}=2.47\times 10^{-3}~\eVq$.  Lower: contours in the plane
    of $\sin^2\theta_{13}$ and the flux normalization
    $f_\text{flux}$. Full regions (lines) correspond to analysis with
    (without) including the RSBL experiments.}
  \label{fig:react-t13}
\end{figure}

The outcome is illustrated in Fig.~\ref{fig:react-t13} where in the
upper panels we show $\Delta\chi^2$ from the Chooz and Palo Verde,
Double Chooz, Daya Bay and Reno experiments as a function of
$\sin^2\theta_{13}$ for various assumptions on the fluxes.  As seen in
the left upper panel, when the new fluxes are taken at face value and
RSBL reactor experiments are not included in the fit, even Chooz and
Palo Verde would prefer $\theta_{13} > 0$ but as soon as RSBL reactor
experiments are included in the fit, the preference essentially
disappears~\cite{Schwetz:2011qt} independently of whether the
normalization of the fluxes is also left as a free parameter.  The
lower left panel shows the contours in the plane of
$\sin^2\theta_{13}$ and the flux normalization $f_\text{flux}$ for the
analysis with free reactor flux normalization with and without
including RSBL data.  Central panels show the dependence of the value
of $\theta_{13}$ obtained from the analysis of Double Chooz.  As seen
in the figure, the best fit value as well as the statistical
significance of the non-zero $\theta_{13}$ at Double Chooz severely
depends on the reactor flux assumption. This is due to the lack of the
near detector in Double Chooz at present.\footnote{Let us mention that
  the official Double Chooz analysis~\cite{Abe:2011fz, Abe:2012uy}
  adopts the Bugey4 measurment of the reactor flux as normalization
  for the reactor flux, and therefore it is similar to our ``Free
  Fluxes + RSBL'' analysis.}  Conversely the right panels show the
results of Daya Bay and Reno which rely on a near-far detector
comparison and therefore are independent of the overall flux
normalization.\footnote{Note that to-date neither Daya Bay nor Reno
  have published results on the absolute flux determination. Therefore
  we always have to include free normalization factors in the fit to
  their data.}  Once they are included in the global analysis the
impact of RSBL data is reduced but it still makes a difference in the
final best fit values and ranges. To quantify this effect we chose to
show the results of the global analysis for the two limiting
assumptions of either taking the predicted fluxes (with the related
uncertainties and correlations) of~\cite{Huber:2011wv} and ignore the
RSBL data (which we label in the figures as ``Huber'') or to allow for
a free normalization of the reactor fluxes and include the RSBL data
to reduce its possible allowed range (labeled as ``Free Fluxes +
RSBL'').

\begin{figure}\centering
  \includegraphics[width=0.6\textwidth]{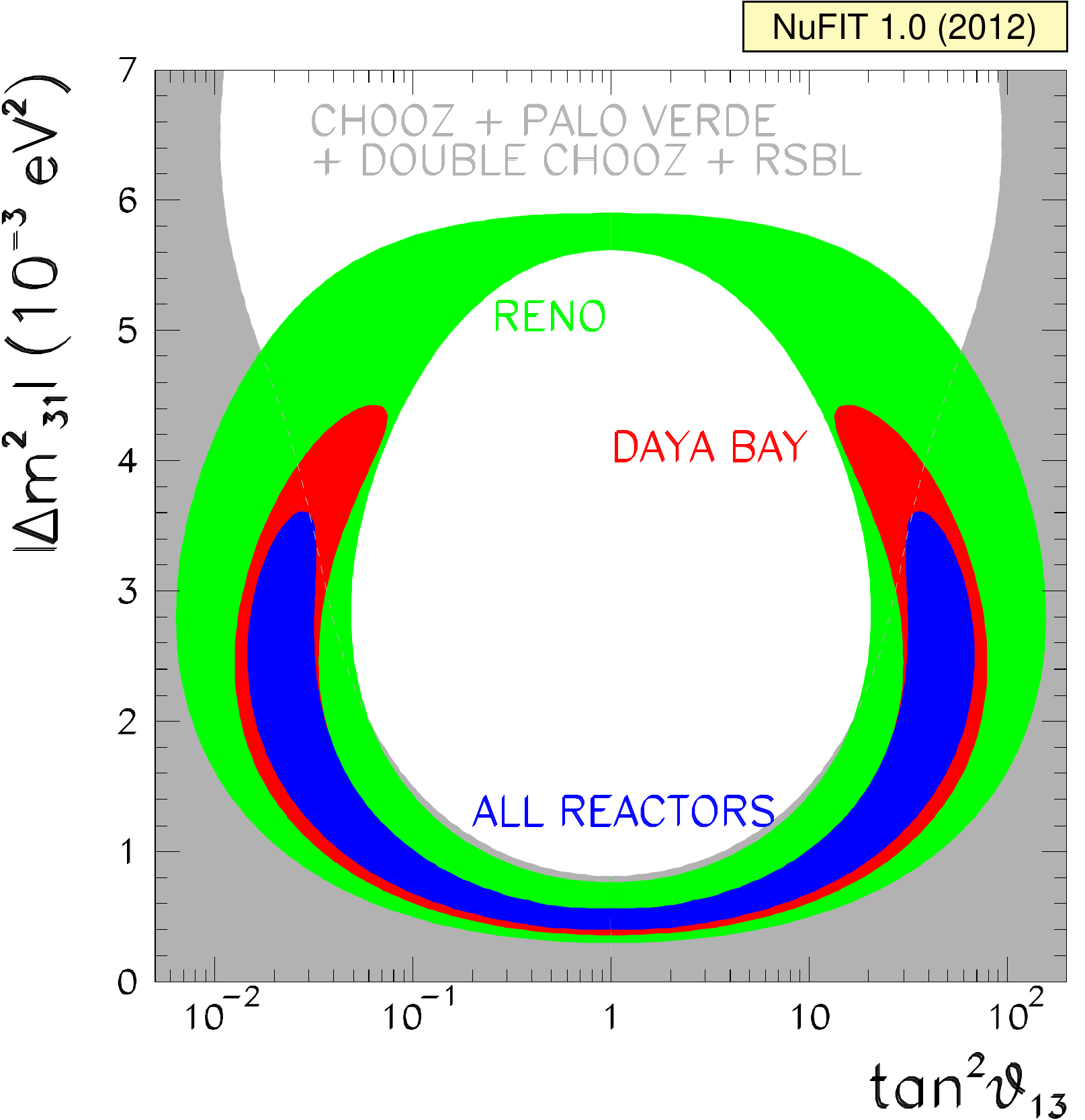}
  \caption{$3\sigma$ allowed regions in the plane of $|\Dmq_{31}|$ and
    $\sin^2\theta_{13}$ for different combinations of the reactor
    experiments.  The region labeled ``ALL REACTORS'' does not include
    Kamland.}
  \label{fig:region-react}
\end{figure}

It is also interesting to notice that since the dominant oscillation
probability in these reactor experiments with $L \sim 1$~km
is\footnote{In the numerical analysis higher order effects associated
  to $\Dmq_{21}$ are included, and they have a noticeable effect on
  the extraction of $\theta_{13}$, especially for Daya Bay.  In
  principle those effects could even distinguish Normal and Inverted
  orderings~\cite{Petcov:2001sy, Learned:2006wy, Ghoshal:2010wt}, an
  effect which however is below the present sensitivity of the
  experiments.}
\begin{equation}
  \label{eq:peereac}
  P_{\nu_e\to\nu_e} = 1 - \sin^22\theta_{13} \sin^2
  \left(\frac{\Dmq_{31} L}{4E} \right)+
  \mathcal{O}(\alpha^2) \,,
\end{equation}
with $\alpha \equiv \Dmq_{21} / \Dmq_{31}$, then the rates observed in
the detectors at different baselines can provide an independent
determination of $\Dmq_{31}$~\cite{Bezerra:2012at}.  We show in
Fig.~\ref{fig:region-react} the $3\sigma$ allowed regions in the plane
$|\Dmq_{31}|$ versus $\sin^2\theta_{13}$ for different combinations of
the reactor experiments. Of course the accuracy on $|\Dmq_{31}|$ from
these data is much worse than the one from MINOS (although
consistent). One may expect improved sensitivity for $|\Dmq_{31}|$
from reactors once spectral data from Daya Bay and Reno will be
included.

\begin{figure}\centering
  \includegraphics[width=0.9\textwidth]{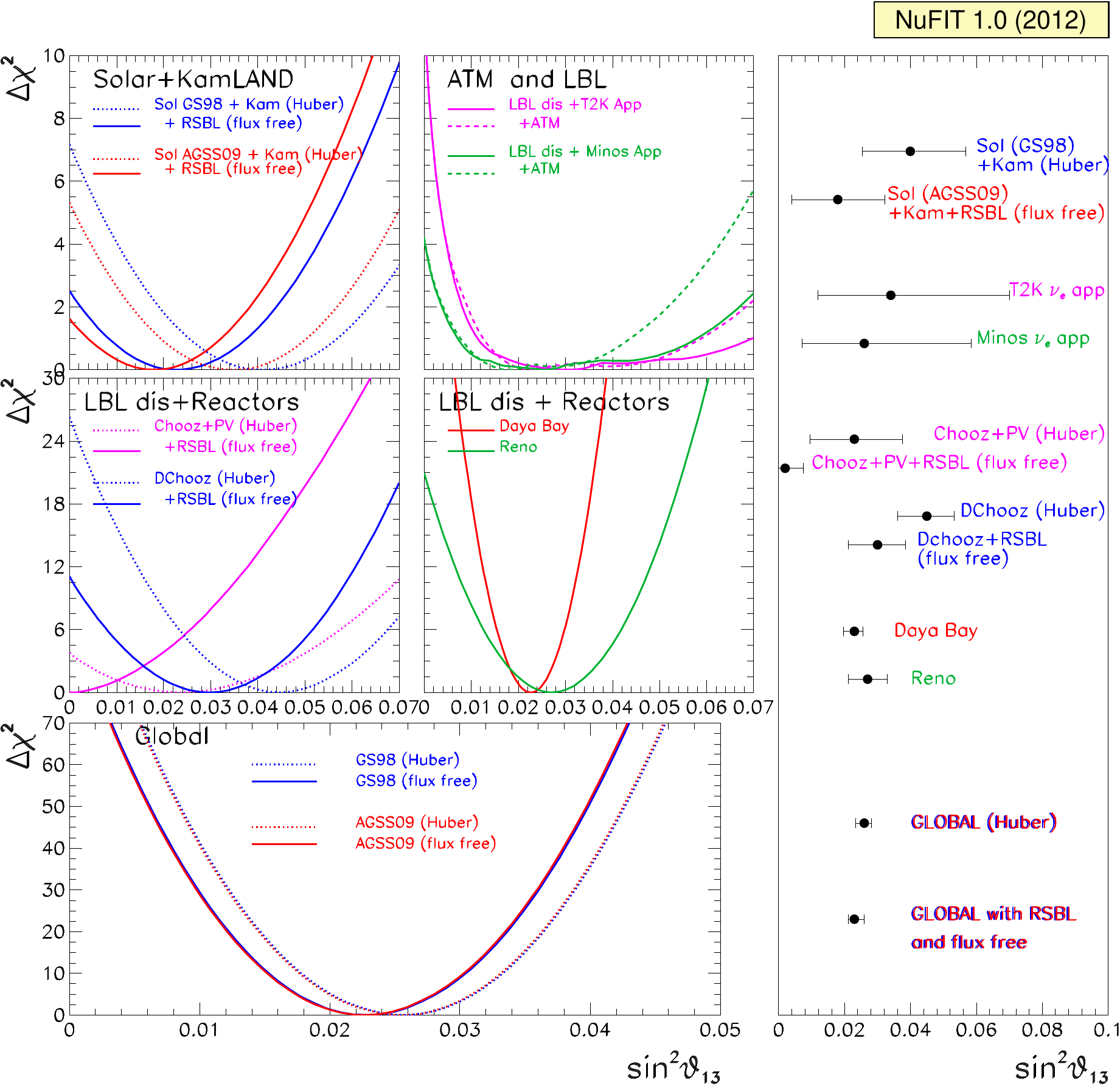}
  \caption{Dependence of $\Delta\chi^2$ on $\sin^2\theta_{13}$ for the
    different data samples and assumptions as labeled in the figure
    and the corresponding $1\sigma$ ranges.}
  \label{fig:chisq-t13}
\end{figure}

We conclude this section by presenting in Fig.~\ref{fig:chisq-t13} the
dependence of $\Delta\chi^2$ on $\sin^2\theta_{13}$ for the different
data samples included in the global analysis. In the upper left panel
we summarize the present status of the ``hint'' for a non-zero
$\theta_{13}$ from the mismatch of the best-fit point values of
$\Dmq_{21}$ and $\theta_{12}$ between the solar analysis and KamLAND in
the framework of 2-$\nu$ oscillations~\cite{Fogli:2008jx,
  Fogli:2009zza, Schwetz:2008er, Maltoni:2008ka, Balantekin:2008zm,
  GonzalezGarcia:2010er}. As discussed in detail in
Ref.~\cite{GonzalezGarcia:2010er}, and seen in the figure, the
statistical significance of this effect depends on the Standard Solar
Model employed in the analysis, either the GS98 model or the
AGSS09~\cite{Serenelli:2009yc}\footnote{GS98 is based on the older
  solar abundances leading to high metallicity and which perfectly
  agreed with helioseismological data. AGSS09 uses the new precise
  determination of the solar abundances which imply a lower
  metallicity and cannot reproduce the helioseismological data. This
  conflict constitutes the so-called ``solar composition problem''.}
and, to lesser extend, on the capture cross-section in gallium
employed\footnote{To explain the lower-than-expected rate observed in
  the calibration of the GALLEX and SAGE detectors (the so-called
  ``Gallium anomaly'') Ref.~\cite{Abdurashitov:2009tn} considers a
  modification of the capture cross-section of
  Ref.~\cite{Bahcall:1997eg} where the contribution from the two
  lowest-lying excited states in \Nuc[71]{Ge} is set to zero.  In our
  analysis labeled ``GS98'' we adopt the cross section
  from~\cite{Bahcall:1997eg}, whereas for the analysis labeled
  ``AGSS09'' we used the modified cross section
  following~\cite{Abdurashitov:2009tn},
  see~\cite{GonzalezGarcia:2010er} for details.}.  As seen in the lower panel of
Fig.~\ref{fig:chisq-t13}, once the solar and KamLAND results are
combined with the oscillation signatures from reactor and LBL
experiments the final allowed range of the oscillation parameters
becomes robust under changes on these factors on the solar analysis.

\section{Determination of $\theta_{23}$ and $\delta_{CP}$}
\label{sec:t23del}

As seen in Sec.~\ref{sec:global} our results show that non-maximal
$\theta_{23}$ is favoured at the level of $\sim 2\sigma$ ($\sim 1.5 \sigma$)
for Normal (Inverted) ordering while the statistical significance of the
effects associated with $\delta_\text{CP}$ is $\leq 1.5 \sigma $ ($\leq 1.75
\sigma$) for Normal (Inverted) ordering. Next we discuss the relative role
of LBL, reactor and atmospheric neutrino data in these results, for similar
discussions see also Refs.~\cite{Fogli:2012ua, Tortola:2012te}.

\subsection{The beam--reactor interplay}
\label{sec:beam-react}

Since the advent of data on $\nu_\mu \to \nu_e$ searches from T2K and
MINOS on the one side, and $\theta_{13}$ reactor experiments
Double Chooz, Daya Bay, and Reno on the other side, the long
anticipated complementarity of beam and reactor
experiments~\cite{Minakata:2002jv, Huber:2003pm} is now a reality.
As shown in Sec.~\ref{sec:global}, the global analysis indicates a
deviation of $\theta_{23}$ from the maximal mixing value of
$45^\circ$, roughly at the level of $1.7\sigma-2\sigma$, see
Fig.~\ref{fig:chisq-glob}. If confirmed, such a deviation would have
profound implications for neutrino mass models based on flavour
symmetries. An important contribution to this effect comes
from recent MINOS data on $\nu_\mu$ disappearance. Neglecting effects
of $\Dmq_{21}$ and the matter effect, the relevant survival
probability in MINOS is given by
\begin{equation}
  \label{eq:Pmm}
  P_{\nu_\mu\to\nu_\mu} = 1 - 4|U_{\mu 3}|^2 (1 - |U_{\mu 3}|^2)
  \sin^2\frac{\Dmq_{31} L}{4E} \,,\qquad |U_{\mu 3}|^2 =
  \sin^2\theta_{23} \cos^2\theta_{13} \,,
\end{equation}
where $L$ is the baseline and $E$ is the neutrino energy. Hence, the
probability is symmetric under $|U_{\mu 3}|^2 \to (1-|U_{\mu 3}|^2)$.
In the two-flavour limit of $\theta_{13} = 0$ this implies that the
data is sensitive only to $\sin^22\theta_{23}$, which for $\theta_{23}
\neq 45^\circ$ leads to a degeneracy between the first and second
octants of $\theta_{23}$~\cite{Fogli:1996pv}. Indeed, recent data from
MINOS~\cite{minos:nu2012} give a best fit point of $\sin^22\theta
\approx 0.94$ if analyzed in a two-flavour framework.

Since $\theta_{13}$ is large one can try to explore a synergy between
long-baseline appearance experiments and an independent determination
of $\theta_{13}$ at reactor experiments in order to resolve the
degeneracy~\cite{Fogli:1996pv, Minakata:2002jv, McConnel:2004bd}.  Let
us look at the appearance probability relevant for the
$\nu_\mu\to\nu_e$ searches at T2K and MINOS. Expanding to second order
in the small parameters $\sin\theta_{13}$ and $\alpha \equiv \Dmq_{21}
/ \Dmq_{31}$ and assuming a constant matter density one
finds~\cite{Cervera:2000kp, Freund:2001pn, Akhmedov:2004ny}:
\begin{equation}
  \label{eq:Pme}
  \begin{split}
    P_{\nu_\mu\to\nu_e}
    &\approx 4 \, \sin^2\theta_{13} \, \sin^2\theta_{23}
    \frac{\sin^2 \Delta (1 - A)}{(1 - A)^2} +
    \alpha^2 \sin^2 2\theta_{12} \, \cos^2\theta_{23}
    \frac{\sin^2 A\Delta}{A^2}
    \\
    &+ 2 \, \alpha \, \sin\theta_{13} \, \sin 2\theta_{12} \,
    \sin2\theta_{23} \cos(\Delta \pm \delta_\text{CP}) \,
    \frac{\sin\Delta A}{A} \, \frac{\sin \Delta (1 - A)}{1-A} \,,
  \end{split}
\end{equation}
with the definitions
\begin{equation}
  \Delta \equiv \frac{\Dmq_{31} L}{4E} \,,
  \quad A \equiv \frac{2EV}{\Dmq_{31}} \,,
\end{equation}
where $L$ is the baseline, $E$ is the neutrino energy, and $V$ is the
effective matter potential~\cite{Wolfenstein:1977ue}. Note that
$\alpha$, $\Delta$, and $A$ are sensitive to the sign of $\Dmq_{31}$
(\textit{i.e.}, the type of the neutrino mass ordering). The plus
(minus) sign in Eq.~\eqref{eq:Pme} applies for neutrinos
(antineutrinos), and for antineutrinos $V\to -V$, which implies $A\to
-A$. It is clear from Eq.~\eqref{eq:Pme} that in the case of large
matter effect, $A \gtrsim 1$, the terms $(1 - A)$ depend strongly on
the type of the mass ordering, and for $A = 1$ (possible for neutrinos
and NO, or anti-neutrinos and IO) a resonance is
encountered~\cite{Mikheev:1986gs}.  Numerically one finds for a
typical matter density of $3~\text{g} / \text{cm}^3$
\begin{equation}\label{eq:Amat}
  |A| \simeq 0.09 \,
  \left(\frac{E}{\text{GeV}}\right)
  \left(\frac{|\Dmq_{31}|}{2.5\times 10^{-3}~\eVq}\right)^{-1} \,.
\end{equation}
Since for T2K $E \sim 0.7$~GeV, matter effects are of order few
percent, whereas in experiments like NOvA~\cite{Ayres:2004js} with
$E\sim 2$~GeV we can have $|A| \sim 0.2$. Note that $\alpha^2 \approx
10^{-3}$, which implies that the second term in the first line of
Eq.~\eqref{eq:Pme} gives a very small contribution compared to the
other terms.  An important observation is that the first term in
Eq.~\eqref{eq:Pme} (which dominates for large $\theta_{13}$) depends
on $\sin^2\theta_{23}$ and therefore is sensitive to the octant.
Reactor experiments with $L \sim 1$~km, on the other hand, provide a
measurement of $\theta_{13}$ independent of $\theta_{23}$, see
Eq.~\eqref{eq:peereac}.  Hence, by combining the data from reactor
experiments such as Double Chooz, Daya Bay, and Reno with the
appearance data from T2K and MINOS one should be sensitive in
principle to the octant of $\theta_{23}$.

\begin{figure}\centering
  \includegraphics[width=0.75\textwidth]{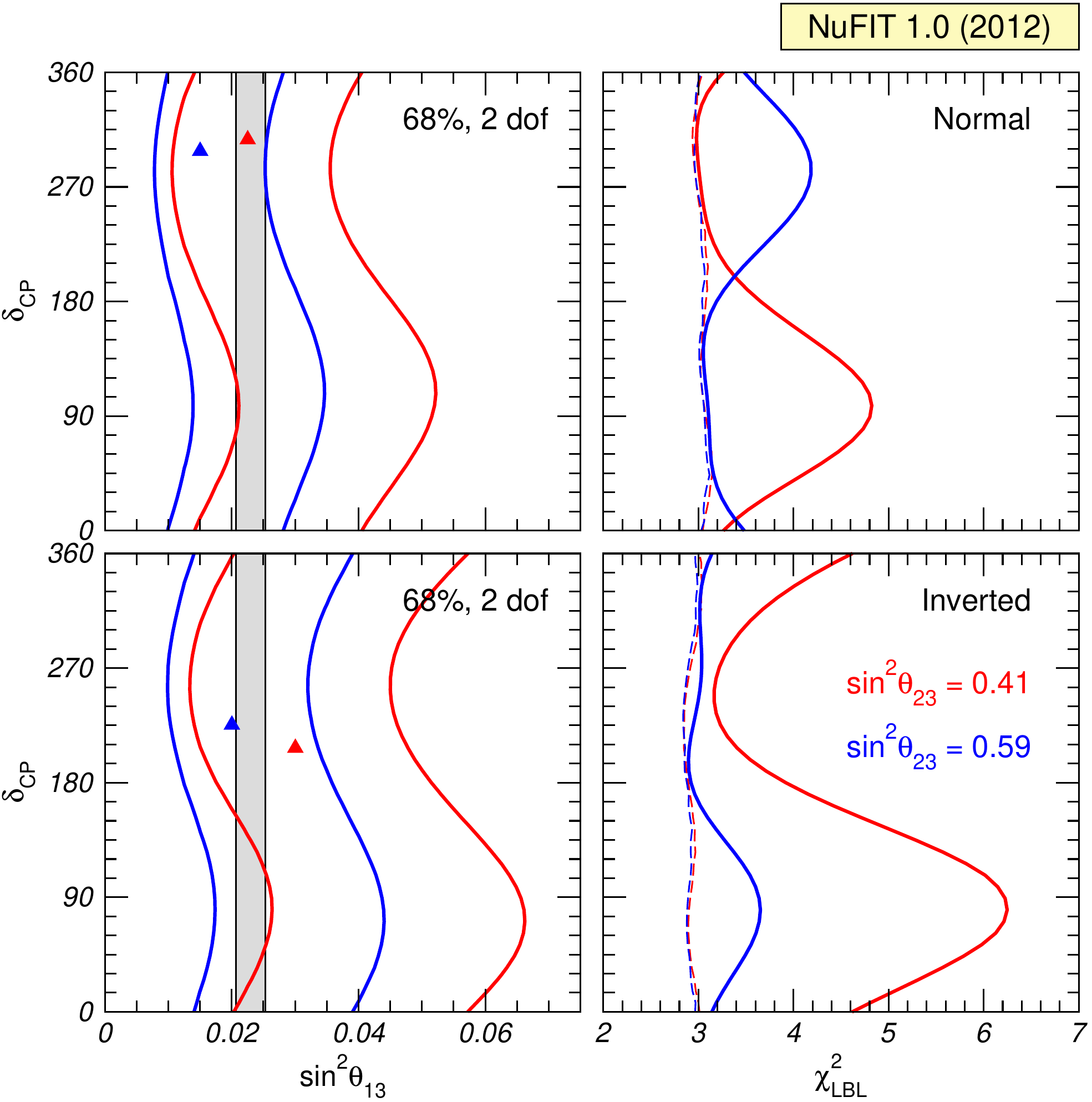}
  \caption{Left: Preferred regions at 68\% CL in the
    $\sin^2\theta_{13}-\delta_{\rm CP}$ plane.  The contour curves
    correspond to T2K + MINOS appearance data, where
    $\sin^2\theta_{23}$ is fixed to the two degenerate solution in the
    1st (red) and 2nd (blue) octant. We define contours for 2~dof with
    respect to the global minimum which is indicated by a triangle.
    The gray region corresponds to the $\theta_{13}$ determination
    from the reactors Double Chooz, Daya Bay, Reno ($1\sigma$ band for
    $\sin^2\theta_{13}$, 1~dof). Right: $\chi^2(\delta)$ from beams
    (dashed) and beams+reactors (solid) with the same color coding as
    in the left panels. The solid curves are computed by adding
    $\Delta\chi^2_{\theta_{13}} = (\sin^2\theta_{13} - 0.023)^2 /
    (0.0023)^2$ to the $\chi^2$ from T2K and MINOS appearance data.
    Upper (lower) panels are for NO (IO). The other oscillation
    parameters are fixed to the best fit values from
    Tab.~\ref{tab:results} (Free Fluxes + RSBL).}
  \label{fig:octant}
\end{figure}

The situation from present data is illustrated in Fig.~\ref{fig:octant},
where we show the determination of $\theta_{13}$ from the beam experiments
T2K and MINOS as a function of the CP phase $\delta_\text{CP}$ and the
octant of $\theta_{23}$, where we have chosen values motivated by the MINOS
disappearance result. The resulting regions in $\sin^2\theta_{13}$ are
compared to the reactor measurements from Double Chooz, Daya Bay, and Reno.
We find that for present data from beams and reactors it is not possible to
distinguish between 1st and 2nd $\theta_{23}$ octants. For both
possibilities overlap regions between beams and reactors can be found,
although at different values of $\delta_\text{CP}$. Therefore, current data
from reactor and long-baseline beam experiments are not able to resolve the
$\theta_{23}$ octant degeneracy.\footnote{Other analyses
(e.g.~\cite{Fogli:2012ua, Tortola:2012te, JThomas})
get results favouring at $\mathcal{O}(1\sigma)$ either one
or the other octant which confirms that the results obtained at this CL
depend on the details of the analysis and are not conclusive.} The lifting
of the degeneracy (at low CL) visible in Fig.~\ref{fig:chisq-glob} appears
due to atmospheric neutrino data, to be discussed below.

In principle the reactor--beam combination should also offer some
sensitivity to the CP phase $\delta_\text{CP}$. This is shown in the
right panels of Fig.~\ref{fig:octant}. We see that if the octant of
$\theta_{23}$ and the neutrino mass ordering were known, already
present data from the beam and reactor experiments used in that figure
would show quite sizeable dependence on the CP phase, depending on
which of the 4 degenerate solutions is considered. However, it is also
clear from the figure that once we marginalize over those four
solutions, $\chi^2(\delta_\text{CP})$ becomes very flat and
essentially all values of $\delta_\text{CP}$ would be consistent
within $\Delta\chi^2 \lesssim 0.5$. This is a real-life example of how
degeneracies can seriously spoil the sensitivity of long-baseline
data~\cite{Barger:2001yr}. The somewhat larger $\delta_\text{CP}$
dependence visible in Fig.~\ref{fig:chisq-glob} follows again from the
global fit including atmospheric neutrinos, as discussed next.

\subsection{The impact of atmospheric neutrinos}
\label{sec:atm}

Atmospheric neutrinos provide a powerful tool to study neutrino
oscillations. The neutrino source contains $\nu_e$ and $\nu_\mu$ as
well as neutrinos and antineutrinos and furthermore, for a good
fraction of the events, neutrinos travel through the Earth matter.  In
the context of 3$\nu$ mixing, the dominant oscillation channel of
atmospheric neutrinos is $\nu_\mu\rightarrow \nu_\tau$ driven by
$|\Dmq_{31}|$ with an amplitude controlled by $\theta_{23}$.  However,
the richness of the atmospheric neutrino beams and baselines opens up
the possibility of sensitivity to subleading oscillation modes,
triggered by $\Dmq_{21}$ and/or $\theta_{13}$, especially in the light
of the new determination of $\theta_{13}$. In particular, they can
shed light on the octant of $\theta_{23}$ and perhaps on the value of
$\delta_\text{CP}$ and the type of neutrino mass ordering.

An interesting observable is the excess of $e$-like events (relative
to the no-oscillation prediction $N_e^0$), since in the two-flavour
limit one expects $N_e = N_e^0$ and therefore any deviation of the
observed number of events from $N_e^0$ should be due to subleading
effects. Such excess can be written in the following way (see,
\textit{e.g.},~\cite{Peres:2003wd}):
\begin{equation}
  \label{eq:PatmNe}
  \begin{split}
    \frac{N_e}{N_e^0} - 1
    & \approx (r \sin^2\theta_{23} - 1) P_{2\nu}(\Dmq_{31}, \theta_{13})
    \\
    & + (r \cos^2\theta_{23} - 1) P_{2\nu}(\Dmq_{21}, \theta_{12})
    \\[2mm]
    & - \sin\theta_{13} \sin 2\theta_{23} \, r \, \Re(A_{ee}^* A_{\mu e}) \,.
  \end{split}
\end{equation}
Here $r \equiv \Phi_\mu / \Phi_e$ is the flux ratio with $r \approx 2$
in the sub-GeV range and $r \approx 2.6 \to 4.5$ in the multi-GeV
range.  $P_{2\nu}(\Dmq, \theta)$ is an effective two-flavour
oscillation probability and $A_{ee}, A_{\mu e}$ are elements of a
transition amplitude matrix. The three terms appearing in
Eq.~\eqref{eq:PatmNe} have a well defined physical interpretation. The
first term is important in the multi-GeV range and is controlled by
the mixing angle $\theta_{13}$ in $P_{2\nu}(\Dmq_{31},
\theta_{13})$. This probability can be strongly affected by resonant
matter effects~\cite{Petcov:1998su, Akhmedov:1998xq, Akhmedov:1998ui,
  Chizhov:1998ug, Chizhov:1999az, Akhmedov:2006hb}. Depending on the
mass hierarchy the resonance will occur either for neutrinos or
antineutrinos.  The second term is important for sub-GeV events and it
takes into account the effect of ``solar oscillations'' due to
$\Dmq_{21}$ and $\theta_{12}$~\cite{Kim:1998bv, Peres:1999yi,
  GonzalezGarcia:2004cu, Akhmedov:2008qt}. Via the pre-factor
containing the flux ratio $r$ both, the first and second terms in
Eq.~\eqref{eq:PatmNe} depend on the octant of $\theta_{23}$, though in
opposite directions: the multi-GeV (sub-GeV) excess is suppressed
(enhanced) for $\theta_{23} < 45^\circ$. Finally, the last term in
Eq.~\eqref{eq:PatmNe} is an interference term between $\theta_{13}$
and $\Dmq_{21}$ amplitudes and this term shows also dependence on the
CP phase $\delta_\text{CP}$~\cite{Peres:2003wd, Akhmedov:2008qt}.

Subdominant three neutrino effects can also affect $\mu$-like events.
For example for multi-GeV muon events one can write the excess in
$\mu$-like events as~\cite{Bernabeu:2003yp, Petcov:2005rv}
\begin{equation}
  \label{eq:PatmNm}
  \begin{split}
    \frac{N_\mu}{N_\mu^0} - 1
    & \approx \sin^2\theta_{23} \left(\frac{1}{r} - \sin^2\theta_{23} \right)
    P_{2\nu}(\Dmq_{31}, \theta_{13})
    \\
    &- \frac{1}{2}\sin^22\theta_{23}\left[1 - \Re(A_{33})\right] \,.
  \end{split}
\end{equation}
The first term is controlled by $\theta_{13}$ and is subject to
resonant matter effects, similar to the first term in
Eq.~\eqref{eq:PatmNe}, though with a different dependence on
$\theta_{23}$ and the flux ratio. In the second term, $A_{33}$ is a
probability amplitude satisfying $P_{2\nu}(\Dmq_{31}, \theta_{13}) = 1
- |A_{33}|^2$. In the limit $\theta_{13} = 0$ we have ${\rm
  Re}(A_{33}) = \cos(\Dmq_{31} L / 2 E)$, such that the second term in
Eq.~\eqref{eq:PatmNm} just describes two-flavour $\nu_\mu\to\nu_\mu$
vacuum oscillations.

\begin{figure}\centering
  \includegraphics[width=0.75\textwidth]{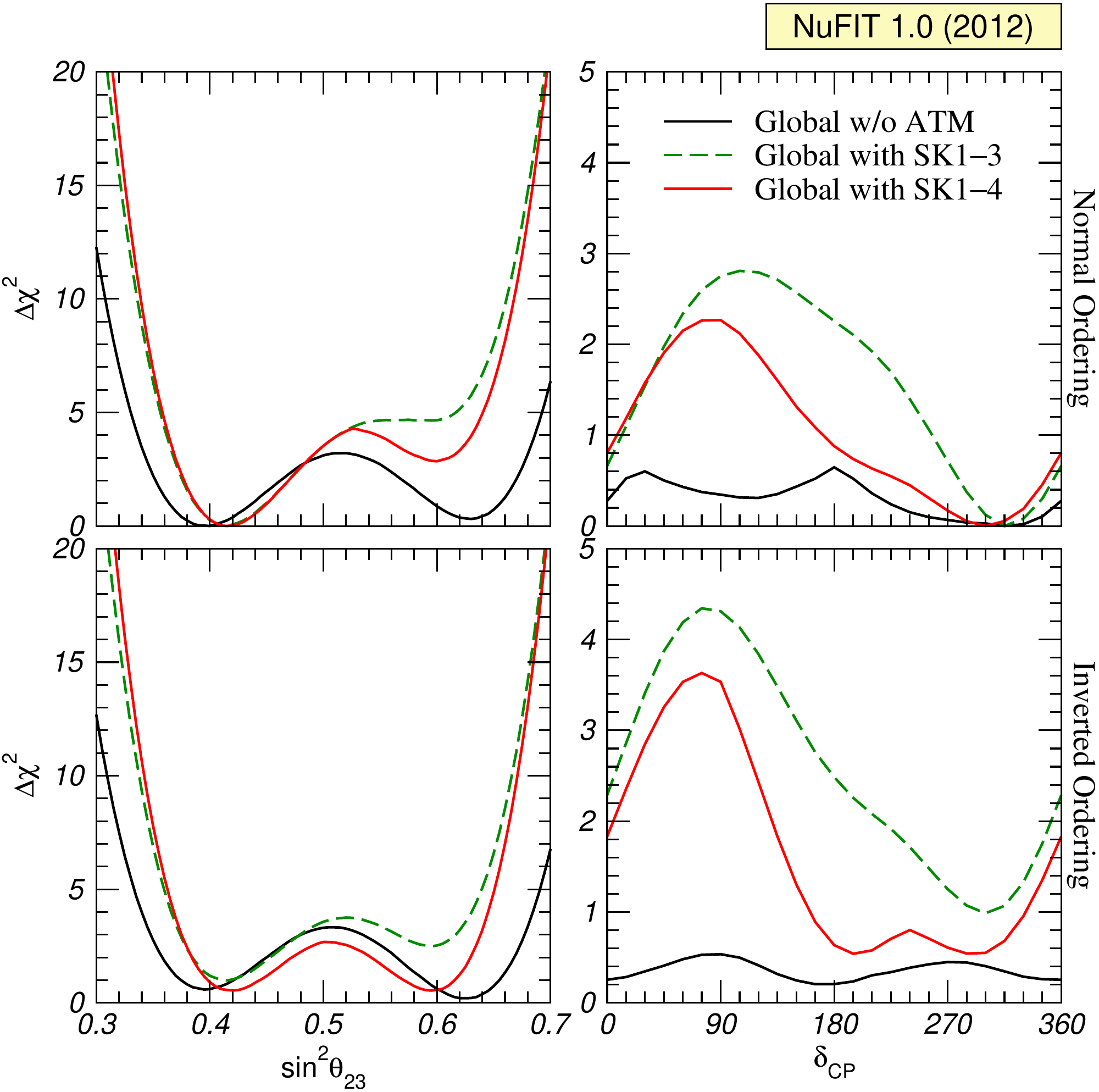}
  \caption{ $\Delta\chi^2$ as a function $\sin^2\theta_{23}$ and
    $\delta_\text{CP}$ for three different analyses as labeled in the
    figure. Upper (lower) panels correspond to Normal (Inverted)
    ordering.}
  \label{fig:chisq-atmos}
\end{figure}

The statistical significance of these effects in the present global
analysis is shown in Fig.~\ref{fig:chisq-atmos} where we show the
dependence of $\Delta\chi^2$ on $\sin^2\theta_{23}$ and
$\delta_\text{CP}$.  The three curves in each panel correspond to the
global analysis including atmospheric data from phases SK1--4 (red
full lines), the analysis without including the atmospheric neutrino
data (black full lines), and the global analysis with the previous
atmospheric sample from phases SK1--3 (green dashed lines). For the
sake of comparison with the results of the analysis of the
Super-Kamiokande collaboration, in our atmospheric analysis with
phases SK1--4 the no-oscillation expectations have been normalized to
those obtained with the new Honda fluxes~\cite{Honda:2011nf}, while
the analysis from SK1--3 is done with the previous set of fluxes
from the same group~\cite{Honda:2006qj}.  For simplicity we only show
the results of the analysis with the normalization of reactor fluxes
is free and data from short-baseline reactor experiments included
(Free Fluxes + RSBL). Similar behaviour is found for the analysis
using reactor fluxes as predicted in~\cite{Huber:2011wv} (Huber).

The figure illustrates that in the analysis without atmospheric
neutrinos (full black lines) a preference for non-maximal value of
$\theta_{23}$ at the level of $1.7\sigma$ ($2\sigma$) for Normal
(Inverted) ordering is observed.  Such result is mainly driven by the
new MINOS $\nu_\mu$ disappearance data~\cite{minos:nu2012}. On the
other hand, we do not observe any statistically relevant sensitivity
to the octant of $\theta_{23}$ from the analysis without atmospheric
neutrinos. This is, as mentioned in the previous section, with the
present data from beams and reactors it is not possible to distinguish
between first and second $\theta_{23}$ octants.

Comparing black with either red or green lines in
Fig.~\ref{fig:chisq-atmos} we see that in the global analysis
including atmospheric data a preference for the first octant is
observed.  This can be attributed to a zenith-angle independent event
excess in the sub-GeV $e$-like data in Super-Kamiokande.  Such an
excess can be explained by oscillations due to
$\Dmq_{21}$~\cite{Kim:1998bv, Peres:1999yi, GonzalezGarcia:2004cu}.
For sub-GeV events the second term in Eq.~\eqref{eq:PatmNe} is
relevant.  In that energy regime $r \approx 2$ and for
$\sin^2\theta_{23} \approx 0.5$ the pre-factor $(r \cos^2\theta_{23} -
1)$ is suppressed, whereas in the first octant with $\sin^2\theta_{23}
< 0.5$ an excess is induced. We also see that despite the larger
statistics, the effect is slightly less significant in the analysis
including the SK4 sample (full red lines) versus the one without that
sample (dashed green lines). This is in agreement with the preliminary
analysis performed by the Super-Kamiokande~\cite{skatm:nu2012}
collaboration of their full phases SK1--4 in which this excess is less
significant compared to phases SK1--3. Let us also mention that due to
the way the data are presented we have to use different binning in
fitting SK1--3 compared to SK1--4, in particular of the sub-GeV
samples, which also affects the sensitivity to the $\theta_{23}$
octant.

Given the preference for the first $\theta_{23}$ octant once
atmospheric data is included, also the sensitivity to
$\delta_\text{CP}$ is somewhat increased as seen in the lower panels
of Fig.~\ref{fig:chisq-atmos}.  This can be understood from both the
effect of $\delta_\text{CP}$ in the atmospheric data, as well as in
the long-baseline experiments.  Looking at Fig.~\ref{fig:octant} we
see that once the solution with $\sin^2\theta_{23} < 45^\circ$ is
favoured, the beam--reactor combination provides a better sensitivity
to $\delta_\text{CP}$, visible in the right panels. This results in
the final sensitivity shown in Fig.~\ref{fig:chisq-atmos}, which is at
the level of $\Delta\chi^2 \approx 3$.  We emphasize again the crucial
interplay of different data sets necessary for this sensitivity to
emerge: MINOS $\nu_\mu$ disappearance prefers $\sin^22\theta_{23} <
1$, atmospheric data disfavours $\sin^2\theta_{23} > 0.5$, and the
atmospheric data together with the $\nu_\mu\to\nu_e$ data from beams
combined with the $\theta_{13}$ determination from reactors provide
sensitivity to $\delta_\text{CP}$.  In brief, in the present global
analysis both the sensitivity to the octant of $\theta_{23}$ and to
$\delta_\text{CP}$ still relies on the observability of sub-dominant
oscillation effects in the atmospheric neutrino analysis.

In this respect it is important to stress that already since SK2 the
Super-Kamiokande collaboration has been presenting its experimental
results in terms of a large number of data samples.  The rates for
some of those samples cannot be theoretically predicted (and therefore
included in a statistical analysis) without a detailed simulation of
the detector, which can only be made by the experimental collaboration
itself. Hence, although our results represent the most up-to-date
analysis of the atmospheric neutrino data which can be performed
outside the collaboration, such an analysis has unavoidable
limitations.\footnote{For details on our simulation of the data
  samples and the statistical analysis see the Appendix of
  Ref.~\cite{GonzalezGarcia:2007ib}.}  For example, as can be seen
from Eqs.~\eqref{eq:PatmNe} and~\eqref{eq:PatmNm} there can be some
features in $e$-like or $\mu$-like data samples which exhibit a
different dependence on $\theta_{23}$, and which of those subtle
effects dominates depends on details of the detector simulation,
binning, and treatment of systematic uncertainties.

\section{Discussion}
\label{sec:dis}

We have presented the results of an updated analysis of solar,
atmospheric, reactor and accelerator neutrino data in the framework of
three-neutrino oscillations.  Quantitatively the results of the
present determination of the oscillation parameters is listed in
Table~\ref{tab:results}. The corresponding leptonic mixing matrix is
given in Eq.~\eqref{eq:umatrix}.

At present the most important players in the new determination of
$\theta_{13}$ are the reactor experiments (see
Fig.~\ref{fig:chisq-t13}).  The global fit excludes $\theta_{13}$
being zero with $\Delta\chi^2 \approx 100$. The results of reactor
experiments without a near detector, in particular Double Chooz, Chooz
and Palo Verde, depend on the expected rates as computed with some
prediction for the neutrino fluxes from the reactors.  This brings up
the possible uncertainty of the determination of $\theta_{13}$
associated with the tension between the recent reevaluation of the
reactor neutrino fluxes in Refs.~\cite{Mueller:2011nm, Huber:2011wv}
and data from reactor experiments with baselines less than 100~m. In
Sec.~\ref{sec:t13} we have quantified this effect and conclude that an
uncertainty on the determination of $\theta_{13}$ at the level of
$1\sigma$ remains due to this so-called reactor
anomaly~\cite{Mention:2011rk}. As the statistics of Daya Bay and Reno
experiments increases and with the operation of the near detector in
Double Chooz this contribution to the error is expected to decrease.

Sec.~\ref{sec:t23del} focuses on our present knowledge of
$\theta_{23}$ and $\delta_\text{CP}$ with special emphasis on the
present limitations of the statistical significance of the possible
deviation of $\theta_{23}$ from maximal, the determination of its
octant and the sensitivity to the CP violating phase. In particular we
have studied the role played in the present analysis by LBL, reactor
and atmospheric neutrinos on these issues.  We find that the global
analysis prefers a non-maximal value of $\theta_{23}$ at the level of
$1.7$--$2 \sigma$ and that this result is mostly driven by the recent
MINOS $\nu_\mu$ disappearance data. However we do not observe any
sensitivity to the octant of $\theta_{23}$ from the global analysis
without atmospheric neutrinos. Although the combination of LBL and
reactor results has the potential to disentangle the octant of
$\theta_{23}$, within their present precision this effect is not
statistically significant as illustrated in Fig.~\ref{fig:octant}.
The lifting of the octant degeneracy (which we find to be at most a
$1.5 \sigma$ effect) visible in Fig.~\ref{fig:chisq-glob} appears due
to atmospheric neutrino data (see also Fig.~\ref{fig:chisq-atmos}).

Equivalently, in principle the reactor--beam combination could also
offer some sensitivity to the CP phase. However, we find that this
effect is not significant enough and without inclusion of atmospheric
neutrinos in the analysis all values of $\delta_\text{CP}$ are
consistent within $\Delta\chi^2 \lesssim 0.5$
(Fig.~\ref{fig:chisq-atmos}). The observed sensitivity to
$\delta_\text{CP}$ in the global analysis is a combined effect of (i)
MINOS $\nu_\mu$ disappearance favouring $\sin^22\theta_{23} < 1$, (ii)
atmospheric data disfavouring $\sin^2\theta_{23} > 0.5$, and (iii) the
$\nu_\mu\to\nu_e$ data from LBL experiments combined with the
$\theta_{13}$ from reactors.  Altogheter they provide a statistical
significance for $\delta_\text{CP}$ at the level of $1.7\sigma$.

We conclude that within the present accuracy of the LBL experiments
and reactor experiments, the small statistical significance of the
determination of the octant of $\theta_{23}$ and of $\delta_\text{CP}$
is driven by the observation of their subdominant effects on the
atmospheric neutrino events.  Finally we mention that, both neutrino
mass orderings (Normal and Inverted) provide a fit of very similar
quality to the global data, with $\Delta\chi^2 \approx 0.5$.

Future updates of this analysis will be provided at the
website Ref.~\cite{nufit}.

\section*{Acknowledgments}

This work is supported by Spanish MINECO (grants FPA2010-20807,
FPA-2009-08958, FPA-2009-09017, FPA2012-34694, consolider-ingenio 2010
grant CSD-2008-0037 and ``Centro de Excelencia Severo Ochoa'' program
SEV-2012-0249), by CUR Generalitat de Catalunya (grant 2009SGR502), by
Comunidad Autonoma de Madrid (HEPHACOS project S2009/ESP-1473), by
USA-NSF (grant PHY-0653342) and by the European Union (EURONU project
FP7-212372 and FP7 Marie Curie-ITN actions PITN-GA-2009-237920
``UNILHC'' and PITN-GA-2011-289442 ``INVISIBLES'').

\bibliographystyle{JHEP}
\bibliography{references}

\providecommand{\href}[2]{#2}\begingroup\raggedright\begin{thebibliography}{10}

\bibitem{Pontecorvo:1967fh}
B.~Pontecorvo, {\it {Neutrino experiments and the question of leptonic-charge
  conservation}},  {\em Sov. Phys. JETP} {\bf 26} (1968) 984--988.

\bibitem{Gribov:1968kq}
V.~N. Gribov and B.~Pontecorvo, {\it {Neutrino astronomy and lepton charge}},
  {\em Phys. Lett.} {\bf B28} (1969) 493.

\bibitem{GonzalezGarcia:2007ib}
M.~C. Gonzalez-Garcia and M.~Maltoni, {\it {Phenomenology with Massive
  Neutrinos}},  {\em Phys. Rept.} {\bf 460} (2008) 1--129,
  [\href{http://xxx.lanl.gov/abs/0704.1800}{{\tt arXiv:0704.1800}}].

\bibitem{Maki:1962mu}
Z.~Maki, M.~Nakagawa, and S.~Sakata, {\it {Remarks on the unified model of
  elementary particles}},  {\em Prog. Theor. Phys.} {\bf 28} (1962) 870--880.

\bibitem{Kobayashi:1973fv}
M.~Kobayashi and T.~Maskawa, {\it {CP Violation in the Renormalizable Theory of
  Weak Interaction}},  {\em Prog. Theor. Phys.} {\bf 49} (1973) 652--657.

\bibitem{Bilenky:1980cx}
S.~M. Bilenky, J.~Hosek, and S.~T. Petcov, {\it {On Oscillations of Neutrinos
  with Dirac and Majorana Masses}},  {\em Phys. Lett.} {\bf B94} (1980) 495.

\bibitem{Langacker:1986jv}
P.~Langacker, S.~T. Petcov, G.~Steigman, and S.~Toshev, {\it {On the
  Mikheev-Smirnov-Wolfenstein (MSW) Mechanism of Amplification of Neutrino
  Oscillations in Matter}},  {\em Nucl. Phys.} {\bf B282} (1987) 589.

\bibitem{An:2012eh}
{\bf DAYA-BAY} Collaboration, F.~An et~al., {\it {Observation of
  electron-antineutrino disappearance at Daya Bay}},  {\em Phys.Rev.Lett.} {\bf
  108} (2012) 171803, [\href{http://xxx.lanl.gov/abs/1203.1669}{{\tt
  arXiv:1203.1669}}].

\bibitem{Ahn:2012nd}
{\bf RENO} Collaboration, J.~Ahn et~al., {\it {Observation of Reactor Electron
  Antineutrino Disappearance in the RENO Experiment}},  {\em Phys.Rev.Lett.}
  {\bf 108} (2012) 191802, [\href{http://xxx.lanl.gov/abs/1204.0626}{{\tt
  arXiv:1204.0626}}].

\bibitem{Abe:2011fz}
{\bf Double Chooz} Collaboration, Y.~Abe et~al., {\it {Indication for the
  disappearance of reactor electron antineutrinos in the Double Chooz
  experiment}},  {\em Phys.Rev.Lett.} {\bf 108} (2012) 131801,
  [\href{http://xxx.lanl.gov/abs/1112.6353}{{\tt arXiv:1112.6353}}].

\bibitem{Abe:2011sj}
{\bf T2K} Collaboration, K.~Abe et~al., {\it {Indication of Electron Neutrino
  Appearance from an Accelerator-produced Off-axis Muon Neutrino Beam}},  {\em
  Phys.Rev.Lett.} {\bf 107} (2011) 041801,
  [\href{http://xxx.lanl.gov/abs/1106.2822}{{\tt arXiv:1106.2822}}].

\bibitem{Adamson:2011qu}
{\bf MINOS} Collaboration, P.~Adamson et~al., {\it {Improved search for
  muon-neutrino to electron-neutrino oscillations in MINOS}},  {\em
  Phys.Rev.Lett.} {\bf 107} (2011) 181802,
  [\href{http://xxx.lanl.gov/abs/1108.0015}{{\tt arXiv:1108.0015}}].

\bibitem{nufit}
NuFit webpage, {\tt www.nu-fit.org}.

\bibitem{Fogli:2012ua}
G.~Fogli, E.~Lisi, A.~Marrone, D.~Montanino, A.~Palazzo, et~al., {\it {Global
  analysis of neutrino masses, mixings and phases: entering the era of leptonic
  CP violation searches}},  {\em Phys.Rev.} {\bf D86} (2012) 013012,
  [\href{http://xxx.lanl.gov/abs/1205.5254}{{\tt arXiv:1205.5254}}].

\bibitem{Tortola:2012te}
D.~Forero, M.~Tortola, and J.~Valle, {\it {Global status of neutrino
  oscillation parameters after Neutrino-2012}},
  \href{http://xxx.lanl.gov/abs/1205.4018}{{\tt arXiv:1205.4018}}.

\bibitem{Machado:2011ar}
P.~Machado, H.~Minakata, H.~Nunokawa, and R.~Zukanovich~Funchal, {\it
  {Combining Accelerator and Reactor Measurements of $\theta_{13}$: The First
  Result}},  {\em JHEP} {\bf 1205} (2012) 023,
  [\href{http://xxx.lanl.gov/abs/1111.3330}{{\tt arXiv:1111.3330}}].

\bibitem{Bergstrom:2012yi}
J.~Bergstrom, {\it {Bayesian evidence for non-zero $\theta_{13}$ and
  CP-violation in neutrino oscillations}},  {\em JHEP} {\bf 1208} (2012) 163,
  [\href{http://xxx.lanl.gov/abs/1205.4404}{{\tt arXiv:1205.4404}}].

\bibitem{skatm:nu2012}
Y.~Itow.
\newblock Talk given at the {\it XXV International Conference on Neutrino
  Physics}, Kyoto, Japan, June 3--9, 2012.

\bibitem{Wendell:2010md}
{\bf Kamiokande} Collaboration, S.-.~R. Wendell et~al., {\it {Atmospheric
  neutrino oscillation analysis with sub-leading effects in Super-Kamiokande I,
  II, and III}},  \href{http://xxx.lanl.gov/abs/1002.3471}{{\tt
  arXiv:1002.3471}}.

\bibitem{Ahn:2006zza}
{\bf K2K} Collaboration, M.~H. Ahn et~al., {\it {Measurement of Neutrino
  Oscillation by the K2K Experiment}},  {\em Phys. Rev.} {\bf D74} (2006)
  072003, [\href{http://xxx.lanl.gov/abs/hep-ex/0606032}{{\tt
  hep-ex/0606032}}].

\bibitem{minos:nu2012}
R.~Nichols.
\newblock Talk given at the {\it XXV International Conference on Neutrino
  Physics}, Kyoto, Japan, June 3--9, 2012.

\bibitem{Adamson:2012rm}
{\bf MINOS} Collaboration, P.~Adamson et~al., {\it {An improved measurement of
  muon antineutrino disappearance in MINOS}},  {\em Phys.Rev.Lett.} {\bf 108}
  (2012) 191801, [\href{http://xxx.lanl.gov/abs/1202.2772}{{\tt
  arXiv:1202.2772}}].

\bibitem{t2k:ichep2012}
K.~Sakashita.
\newblock Talk given at the {\it 36th International Conference on High Energy
  Physics}, Melbourne, Australia, July 4--11, 2012.

\bibitem{Abe:2012gx}
{\bf T2K} Collaboration, K.~Abe et~al., {\it {First Muon-Neutrino Disappearance
  Study with an Off-Axis Beam}},  {\em Phys.Rev.} {\bf D85} (2012) 031103,
  [\href{http://xxx.lanl.gov/abs/1201.1386}{{\tt arXiv:1201.1386}}].

\bibitem{t2k:nu2012}
T.~Nakaya.
\newblock Talk given at the {\it XXV International Conference on Neutrino
  Physics}, Kyoto, Japan, June 3--9, 2012.

\bibitem{Apollonio:1999ae}
{\bf CHOOZ} Collaboration, M.~Apollonio et~al., {\it {Limits on Neutrino
  Oscillations from the CHOOZ Experiment}},  {\em Phys. Lett.} {\bf B466}
  (1999) 415--430, [\href{http://xxx.lanl.gov/abs/hep-ex/9907037}{{\tt
  hep-ex/9907037}}].

\bibitem{Piepke:2002ju}
{\bf Palo Verde} Collaboration, A.~Piepke, {\it {Final results from the Palo
  Verde neutrino oscillation experiment}},  {\em Prog.Part.Nucl.Phys.} {\bf 48}
  (2002) 113--121.

\bibitem{Abe:2012uy}
{\bf Double Chooz} Collaboration, Y.~Abe et~al., {\it {Reactor electron
  antineutrino disappearance in the Double Chooz experiment}},
  \href{http://xxx.lanl.gov/abs/1207.6632}{{\tt arXiv:1207.6632}}.

\bibitem{dc:nu2012}
M.~Ishitsuka.
\newblock Talk given at the {\it XXV International Conference on Neutrino
  Physics}, Kyoto, Japan, June 3--9, 2012.

\bibitem{dayabay:nu2012}
D.~Dwyer.
\newblock Talk given at the {\it XXV International Conference on Neutrino
  Physics}, Kyoto, Japan, June 3--9, 2012.

\bibitem{Gando:2010aa}
{\bf KamLAND} Collaboration, A.~Gando et~al., {\it {Constraints on
  $\theta_{13}$ from A Three-Flavor Oscillation Analysis of Reactor
  Antineutrinos at KamLAND}},  {\em Phys.Rev.} {\bf D83} (2011) 052002,
  [\href{http://xxx.lanl.gov/abs/1009.4771}{{\tt arXiv:1009.4771}}].

\bibitem{Cleveland:1998nv}
B.~T. Cleveland et~al., {\it {Measurement of the solar electron neutrino flux
  with the Homestake chlorine detector}},  {\em Astrophys. J.} {\bf 496} (1998)
  505--526.

\bibitem{Kaether:2010ag}
F.~Kaether, W.~Hampel, G.~Heusser, J.~Kiko, and T.~Kirsten, {\it {Reanalysis of
  the GALLEX solar neutrino flux and source experiments}},  {\em Phys.Lett.}
  {\bf B685} (2010) 47--54, [\href{http://xxx.lanl.gov/abs/1001.2731}{{\tt
  arXiv:1001.2731}}].

\bibitem{Abdurashitov:2009tn}
{\bf SAGE} Collaboration, J.~N. Abdurashitov et~al., {\it {Measurement of the
  solar neutrino capture rate with gallium metal. III: Results for the
  2002--2007 data-taking period}},  {\em Phys. Rev.} {\bf C80} (2009) 015807,
  [\href{http://xxx.lanl.gov/abs/0901.2200}{{\tt arXiv:0901.2200}}].

\bibitem{Hosaka:2005um}
{\bf Super-Kamkiokande} Collaboration, J.~Hosaka et~al., {\it {Solar neutrino
  measurements in Super-Kamiokande-I}},  {\em Phys. Rev.} {\bf D73} (2006)
  112001, [\href{http://xxx.lanl.gov/abs/hep-ex/0508053}{{\tt
  hep-ex/0508053}}].

\bibitem{Aharmim:2007nv}
{\bf SNO} Collaboration, B.~Aharmim et~al., {\it {Measurement of the $\nu_e$
  and total B-8 solar neutrino fluxes with the Sudbury Neutrino Observatory
  phase I data set}},  {\em Phys. Rev.} {\bf C75} (2007) 045502,
  [\href{http://xxx.lanl.gov/abs/nucl-ex/0610020}{{\tt nucl-ex/0610020}}].

\bibitem{Aharmim:2005gt}
{\bf SNO} Collaboration, B.~Aharmim et~al., {\it {Electron energy spectra,
  fluxes, and day-night asymmetries of B-8 solar neutrinos from the 391-day
  salt phase SNO data set}},  {\em Phys. Rev.} {\bf C72} (2005) 055502,
  [\href{http://xxx.lanl.gov/abs/nucl-ex/0502021}{{\tt nucl-ex/0502021}}].

\bibitem{Aharmim:2008kc}
{\bf SNO} Collaboration, B.~Aharmim et~al., {\it {An Independent Measurement of
  the Total Active 8B Solar Neutrino Flux Using an Array of 3He Proportional
  Counters at the Sudbury Neutrino Observatory}},  {\em Phys. Rev. Lett.} {\bf
  101} (2008) 111301, [\href{http://xxx.lanl.gov/abs/0806.0989}{{\tt
  arXiv:0806.0989}}].

\bibitem{Aharmim:2011vm}
{\bf SNO} Collaboration, B.~Aharmim et~al., {\it {Combined Analysis of all
  Three Phases of Solar Neutrino Data from the Sudbury Neutrino Observatory}},
  \href{http://xxx.lanl.gov/abs/1109.0763}{{\tt arXiv:1109.0763}}.

\bibitem{Bellini:2011rx}
{\bf Borexino} Collaboration, G.~Bellini et~al., {\it {Precision measurement of
  the 7Be solar neutrino interaction rate in Borexino}},  {\em Phys.Rev.Lett.}
  {\bf 107} (2011) 141302, [\href{http://xxx.lanl.gov/abs/1104.1816}{{\tt
  arXiv:1104.1816}}].

\bibitem{Collaboration:2008mr}
{\bf Borexino} Collaboration, G.~Bellini et~al., {\it {Measurement of the solar
  8B neutrino flux with 246 live days of Borexino and observation of the MSW
  vacuum-matter transition}},  \href{http://xxx.lanl.gov/abs/0808.2868}{{\tt
  arXiv:0808.2868}}.

\bibitem{Huber:2011wv}
P.~Huber, {\it {On the determination of anti-neutrino spectra from nuclear
  reactors}},  {\em Phys.Rev.} {\bf C84} (2011) 024617,
  [\href{http://xxx.lanl.gov/abs/1106.0687}{{\tt arXiv:1106.0687}}].

\bibitem{GonzalezGarcia:2003qf}
M.~Gonzalez-Garcia and C.~Pena-Garay, {\it {Three neutrino mixing after the
  first results from K2K and KamLAND}},  {\em Phys.Rev.} {\bf D68} (2003)
  093003, [\href{http://xxx.lanl.gov/abs/hep-ph/0306001}{{\tt
  hep-ph/0306001}}].

\bibitem{Schreckenbach:1985ep}
K.~Schreckenbach, G.~Colvin, W.~Gelletly, and F.~Von~Feilitzsch, {\it
  Determination of the anti-neutrino spectrum from u-235 thermal neutron
  fission products up to 9.5-mev},  {\em Phys.Lett.} {\bf B160} (1985)
  325--330.

\bibitem{Mueller:2011nm}
T.~Mueller, D.~Lhuillier, M.~Fallot, A.~Letourneau, S.~Cormon, et~al., {\it
  {Improved Predictions of Reactor Antineutrino Spectra}},  {\em Phys.Rev.}
  {\bf C83} (2011) 054615, [\href{http://xxx.lanl.gov/abs/1101.2663}{{\tt
  arXiv:1101.2663}}].

\bibitem{Mention:2011rk}
G.~Mention, M.~Fechner, T.~Lasserre, T.~Mueller, D.~Lhuillier, et~al., {\it
  {The Reactor Antineutrino Anomaly}},  {\em Phys.Rev.} {\bf D83} (2011)
  073006, [\href{http://xxx.lanl.gov/abs/1101.2755}{{\tt arXiv:1101.2755}}].

\bibitem{Schwetz:2011qt}
T.~Schwetz, M.~Tortola, and J.~Valle, {\it {Global neutrino data and recent
  reactor fluxes: status of three-flavour oscillation parameters}},  {\em New
  J.Phys.} {\bf 13} (2011) 063004,
  [\href{http://xxx.lanl.gov/abs/1103.0734}{{\tt arXiv:1103.0734}}].

\bibitem{Ciuffoli:2012yd}
E.~Ciuffoli, J.~Evslin, and H.~Li, {\it {The Reactor Anomaly after Daya Bay and
  RENO}},  \href{http://xxx.lanl.gov/abs/1205.5499}{{\tt arXiv:1205.5499}}.

\bibitem{Declais:1994ma}
Y.~Declais, H.~de~Kerret, B.~Lefievre, M.~Obolensky, A.~Etenko, et~al., {\it
  {Study of reactor anti-neutrino interaction with proton at Bugey nuclear
  power plant}},  {\em Phys.Lett.} {\bf B338} (1994) 383--389.

\bibitem{Kuvshinnikov:1990ry}
A.~Kuvshinnikov, L.~Mikaelyan, S.~Nikolaev, M.~Skorokhvatov, and A.~Etenko,
  {\it {Measuring the $\bar\nu_e + p \to n + e^+$ cross-section and beta decay
  axial constant in a new experiment at Rovno NPP reactor. (In Russian)}},
  {\em JETP Lett.} {\bf 54} (1991) 253--257.

\bibitem{Declais:1994su}
Y.~Declais, J.~Favier, A.~Metref, H.~Pessard, B.~Achkar, et~al., {\it {Search
  for neutrino oscillations at 15-meters, 40-meters, and 95-meters from a
  nuclear power reactor at Bugey}},  {\em Nucl.Phys.} {\bf B434} (1995)
  503--534.

\bibitem{Vidyakin:1987ue}
G.~Vidyakin, V.~Vyrodov, I.~Gurevich, Y.~Kozlov, V.~Martemyanov, et~al., {\it
  Detection of anti-neutrinos in the flux from two reactors},  {\em
  Sov.Phys.JETP} {\bf 66} (1987) 243--247.

\bibitem{Vidyakin:1994ut}
G.~Vidyakin, V.~Vyrodov, Y.~Kozlov, A.~Martemyanov, V.~Martemyanov, et~al.,
  {\it {Limitations on the characteristics of neutrino oscillations}},  {\em
  JETP Lett.} {\bf 59} (1994) 390--393.

\bibitem{Kwon:1981ua}
H.~Kwon, F.~Boehm, A.~Hahn, H.~Henrikson, J.~Vuilleumier, et~al., {\it Search
  for neutrino oscillations at a fission reactor},  {\em Phys.Rev.} {\bf D24}
  (1981) 1097--1111.

\bibitem{Zacek:1986cu}
{\bf CALTECH-SIN-TUM} Collaboration, G.~Zacek et~al., {\it {Neutrino
  Oscillation Experiments at the Gosgen Nuclear Power Reactor}},  {\em
  Phys.Rev.} {\bf D34} (1986) 2621--2636.

\bibitem{Greenwood:1996pb}
Z.~D. Greenwood et~al., {\it {Results of a two position reactor neutrino
  oscillation experiment}},  {\em Phys. Rev.} {\bf D53} (1996) 6054--6064.

\bibitem{Afonin:1988gx}
A.~Afonin, S.~Ketov, V.~Kopeikin, L.~Mikaelyan, M.~Skorokhvatov, et~al., {\it
  {A study of the reaction $\bar\nu_e + p \to e^+ + n$ on a nuclear reactor}},
  {\em Sov.Phys.JETP} {\bf 67} (1988) 213--221.

\bibitem{Petcov:2001sy}
S.~Petcov and M.~Piai, {\it {The LMA MSW solution of the solar neutrino
  problem, inverted neutrino mass hierarchy and reactor neutrino experiments}},
   {\em Phys.Lett.} {\bf B533} (2002) 94--106,
  [\href{http://xxx.lanl.gov/abs/hep-ph/0112074}{{\tt hep-ph/0112074}}].

\bibitem{Learned:2006wy}
J.~Learned, S.~T. Dye, S.~Pakvasa, and R.~C. Svoboda, {\it {Determination of
  neutrino mass hierarchy and theta(13) with a remote detector of reactor
  antineutrinos}},  {\em Phys.Rev.} {\bf D78} (2008) 071302,
  [\href{http://xxx.lanl.gov/abs/hep-ex/0612022}{{\tt hep-ex/0612022}}].

\bibitem{Ghoshal:2010wt}
P.~Ghoshal and S.~Petcov, {\it {Neutrino Mass Hierarchy Determination Using
  Reactor Antineutrinos}},  {\em JHEP} {\bf 1103} (2011) 058,
  [\href{http://xxx.lanl.gov/abs/1011.1646}{{\tt arXiv:1011.1646}}].

\bibitem{Bezerra:2012at}
T.~Bezerra, H.~Furuta, and F.~Suekane, {\it {Measurement of Effective $\Delta
  m_{31}^2$ using Baseline Differences of Daya Bay, RENO and Double Chooz
  Reactor Neutrino Experiments}},
  \href{http://xxx.lanl.gov/abs/1206.6017}{{\tt arXiv:1206.6017}}.

\bibitem{Fogli:2008jx}
G.~L. Fogli, E.~Lisi, A.~Marrone, A.~Palazzo, and A.~M. Rotunno, {\it {Hints of
  $\theta_{13}>0$ from global neutrino data analysis}},  {\em Phys. Rev. Lett.}
  {\bf 101} (2008) 141801, [\href{http://xxx.lanl.gov/abs/0806.2649}{{\tt
  arXiv:0806.2649}}].

\bibitem{Fogli:2009zza}
G.~L. Fogli, E.~Lisi, A.~Marrone, A.~Palazzo, and A.~M. Rotunno, {\it {Neutrino
  masses and mixing: 2008 status}},  {\em Nucl. Phys. Proc. Suppl.} {\bf 188}
  (2009) 27--30.

\bibitem{Schwetz:2008er}
T.~Schwetz, M.~A. Tortola, and J.~W.~F. Valle, {\it {Three-flavour neutrino
  oscillation update}},  {\em New J. Phys.} {\bf 10} (2008) 113011,
  [\href{http://xxx.lanl.gov/abs/0808.2016}{{\tt arXiv:0808.2016}}].

\bibitem{Maltoni:2008ka}
M.~Maltoni and T.~Schwetz, {\it {Three-flavour neutrino oscillation update and
  comments on possible hints for a non-zero $\theta_{13}$}},
  \href{http://xxx.lanl.gov/abs/0812.3161}{{\tt arXiv:0812.3161}}.

\bibitem{Balantekin:2008zm}
A.~B. Balantekin and D.~Yilmaz, {\it {Contrasting solar and reactor neutrinos
  with a non-zero value of theta13}},  {\em J. Phys.} {\bf G35} (2008) 075007,
  [\href{http://xxx.lanl.gov/abs/0804.3345}{{\tt arXiv:0804.3345}}].

\bibitem{GonzalezGarcia:2010er}
M.~Gonzalez-Garcia, M.~Maltoni, and J.~Salvado, {\it {Updated global fit to
  three neutrino mixing: status of the hints of $\theta_{13} > 0$}},  {\em
  JHEP} {\bf 1004} (2010) 056, [\href{http://xxx.lanl.gov/abs/1001.4524}{{\tt
  arXiv:1001.4524}}].

\bibitem{Serenelli:2009yc}
A.~Serenelli, S.~Basu, J.~W. Ferguson, and M.~Asplund, {\it {New Solar
  Composition: The Problem With Solar Models Revisited}},
  \href{http://xxx.lanl.gov/abs/0909.2668}{{\tt arXiv:0909.2668}}.

\bibitem{Bahcall:1997eg}
J.~N. Bahcall, {\it {Gallium solar neutrino experiments: Absorption cross
  sections, neutrino spectra, and predicted event rates}},  {\em Phys. Rev.}
  {\bf C56} (1997) 3391--3409,
  [\href{http://xxx.lanl.gov/abs/hep-ph/9710491}{{\tt hep-ph/9710491}}].

\bibitem{Minakata:2002jv}
H.~Minakata, H.~Sugiyama, O.~Yasuda, K.~Inoue, and F.~Suekane, {\it {Reactor
  measurement of theta(13) and its complementarity to long baseline
  experiments}},  {\em Phys.Rev.} {\bf D68} (2003) 033017,
  [\href{http://xxx.lanl.gov/abs/hep-ph/0211111}{{\tt hep-ph/0211111}}].

\bibitem{Huber:2003pm}
P.~Huber, M.~Lindner, T.~Schwetz, and W.~Winter, {\it {Reactor neutrino
  experiments compared to superbeams}},  {\em Nucl.Phys.} {\bf B665} (2003)
  487--519, [\href{http://xxx.lanl.gov/abs/hep-ph/0303232}{{\tt
  hep-ph/0303232}}].

\bibitem{Fogli:1996pv}
G.~L. Fogli and E.~Lisi, {\it {Tests of three flavor mixing in long baseline
  neutrino oscillation experiments}},  {\em Phys.Rev.} {\bf D54} (1996)
  3667--3670, [\href{http://xxx.lanl.gov/abs/hep-ph/9604415}{{\tt
  hep-ph/9604415}}].

\bibitem{McConnel:2004bd}
K.~B.~M. Mahn and M.~H. Shaevitz, {\it {Comparisons and combinations of reactor
  and long-baseline neutrino oscillation measurements}},  {\em Int.J.Mod.Phys.}
  {\bf A21} (2006) 3825--3844,
  [\href{http://xxx.lanl.gov/abs/hep-ex/0409028}{{\tt hep-ex/0409028}}].

\bibitem{Cervera:2000kp}
A.~Cervera, A.~Donini, M.~Gavela, J.~Gomez~Cadenas, P.~Hernandez, et~al., {\it
  {Golden measurements at a neutrino factory}},  {\em Nucl.Phys.} {\bf B579}
  (2000) 17--55, [\href{http://xxx.lanl.gov/abs/hep-ph/0002108}{{\tt
  hep-ph/0002108}}].

\bibitem{Freund:2001pn}
M.~Freund, {\it {Analytic approximations for three neutrino oscillation
  parameters and probabilities in matter}},  {\em Phys.Rev.} {\bf D64} (2001)
  053003, [\href{http://xxx.lanl.gov/abs/hep-ph/0103300}{{\tt
  hep-ph/0103300}}].

\bibitem{Akhmedov:2004ny}
E.~K. Akhmedov, R.~Johansson, M.~Lindner, T.~Ohlsson, and T.~Schwetz, {\it
  {Series expansions for three flavor neutrino oscillation probabilities in
  matter}},  {\em JHEP} {\bf 0404} (2004) 078,
  [\href{http://xxx.lanl.gov/abs/hep-ph/0402175}{{\tt hep-ph/0402175}}].

\bibitem{Wolfenstein:1977ue}
L.~Wolfenstein, {\it {Neutrino oscillations in matter}},  {\em Phys. Rev.} {\bf
  D17} (1978) 2369--2374.

\bibitem{Mikheev:1986gs}
S.~P. Mikheev and A.~Y. Smirnov, {\it {Resonance enhancement of oscillations in
  matter and solar neutrino spectroscopy}},  {\em Sov. J. Nucl. Phys.} {\bf 42}
  (1985) 913--917.

\bibitem{Ayres:2004js}
{\bf NOvA} Collaboration, D.~Ayres et~al., {\it {NOvA: Proposal to build a 30
  kiloton off-axis detector to study $\nu_\mu \to \nu_e$ oscillations in the
  NuMI beamline}},  \href{http://xxx.lanl.gov/abs/hep-ex/0503053}{{\tt
  hep-ex/0503053}}.

\bibitem{JThomas}
J.~Thomas.
\newblock Talk given at {\it Neutrinos at the forefront of elementary particle
  physics and astrophysics}, Lyon, France, Oct. 22–24, 2012.

\bibitem{Barger:2001yr}
V.~Barger, D.~Marfatia, and K.~Whisnant, {\it {Breaking eight fold degeneracies
  in neutrino CP violation, mixing, and mass hierarchy}},  {\em Phys.Rev.} {\bf
  D65} (2002) 073023, [\href{http://xxx.lanl.gov/abs/hep-ph/0112119}{{\tt
  hep-ph/0112119}}].

\bibitem{Peres:2003wd}
O.~Peres and A.~Y. Smirnov, {\it {Atmospheric neutrinos: LMA oscillations,
  U(e3) induced interference and CP violation}},  {\em Nucl.Phys.} {\bf B680}
  (2004) 479--509, [\href{http://xxx.lanl.gov/abs/hep-ph/0309312}{{\tt
  hep-ph/0309312}}].

\bibitem{Petcov:1998su}
S.~Petcov, {\it {Diffractive - like (or parametric resonance - like?)
  enhancement of the earth (day - night) effect for solar neutrinos crossing
  the earth core}},  {\em Phys.Lett.} {\bf B434} (1998) 321--332,
  [\href{http://xxx.lanl.gov/abs/hep-ph/9805262}{{\tt hep-ph/9805262}}].

\bibitem{Akhmedov:1998xq}
E.~K. Akhmedov, A.~Dighe, P.~Lipari, and A.~Smirnov, {\it {Atmospheric
  neutrinos at Super-Kamiokande and parametric resonance in neutrino
  oscillations}},  {\em Nucl.Phys.} {\bf B542} (1999) 3--30,
  [\href{http://xxx.lanl.gov/abs/hep-ph/9808270}{{\tt hep-ph/9808270}}].

\bibitem{Akhmedov:1998ui}
E.~K. Akhmedov, {\it {Parametric resonance of neutrino oscillations and passage
  of solar and atmospheric neutrinos through the earth}},  {\em Nucl.Phys.}
  {\bf B538} (1999) 25--51, [\href{http://xxx.lanl.gov/abs/hep-ph/9805272}{{\tt
  hep-ph/9805272}}].

\bibitem{Chizhov:1998ug}
M.~Chizhov, M.~Maris, and S.~Petcov, {\it {On the oscillation length resonance
  in the transitions of solar and atmospheric neutrinos crossing the earth
  core}},  \href{http://xxx.lanl.gov/abs/hep-ph/9810501}{{\tt hep-ph/9810501}}.

\bibitem{Chizhov:1999az}
M.~Chizhov and S.~Petcov, {\it {New conditions for a total neutrino conversion
  in a medium}},  {\em Phys.Rev.Lett.} {\bf 83} (1999) 1096--1099,
  [\href{http://xxx.lanl.gov/abs/hep-ph/9903399}{{\tt hep-ph/9903399}}].

\bibitem{Akhmedov:2006hb}
E.~K. Akhmedov, M.~Maltoni, and A.~Y. Smirnov, {\it {1-3 leptonic mixing and
  the neutrino oscillograms of the Earth}},  {\em JHEP} {\bf 0705} (2007) 077,
  [\href{http://xxx.lanl.gov/abs/hep-ph/0612285}{{\tt hep-ph/0612285}}].

\bibitem{Kim:1998bv}
C.~Kim and U.~Lee, {\it {Comment on the possible electron neutrino excess in
  the Super-Kamiokande atmospheric neutrino experiment}},  {\em Phys.Lett.}
  {\bf B444} (1998) 204--207,
  [\href{http://xxx.lanl.gov/abs/hep-ph/9809491}{{\tt hep-ph/9809491}}].

\bibitem{Peres:1999yi}
O.~Peres and A.~Y. Smirnov, {\it {Testing the solar neutrino conversion with
  atmospheric neutrinos}},  {\em Phys.Lett.} {\bf B456} (1999) 204--213,
  [\href{http://xxx.lanl.gov/abs/hep-ph/9902312}{{\tt hep-ph/9902312}}].

\bibitem{GonzalezGarcia:2004cu}
M.~Gonzalez-Garcia, M.~Maltoni, and A.~Y. Smirnov, {\it {Measuring the
  deviation of the 2-3 lepton mixing from maximal with atmospheric neutrinos}},
   {\em Phys.Rev.} {\bf D70} (2004) 093005,
  [\href{http://xxx.lanl.gov/abs/hep-ph/0408170}{{\tt hep-ph/0408170}}].

\bibitem{Akhmedov:2008qt}
E.~K. Akhmedov, M.~Maltoni, and A.~Y. Smirnov, {\it {Neutrino oscillograms of
  the Earth: Effects of 1-2 mixing and CP-violation}},  {\em JHEP} {\bf 0806}
  (2008) 072, [\href{http://xxx.lanl.gov/abs/0804.1466}{{\tt
  arXiv:0804.1466}}].

\bibitem{Bernabeu:2003yp}
J.~Bernabeu, S.~Palomares~Ruiz, and S.~Petcov, {\it {Atmospheric neutrino
  oscillations, theta(13) and neutrino mass hierarchy}},  {\em Nucl.Phys.} {\bf
  B669} (2003) 255--276, [\href{http://xxx.lanl.gov/abs/hep-ph/0305152}{{\tt
  hep-ph/0305152}}].

\bibitem{Petcov:2005rv}
S.~Petcov and T.~Schwetz, {\it {Determining the neutrino mass hierarchy with
  atmospheric neutrinos}},  {\em Nucl.Phys.} {\bf B740} (2006) 1--22,
  [\href{http://xxx.lanl.gov/abs/hep-ph/0511277}{{\tt hep-ph/0511277}}].

\bibitem{Honda:2011nf}
M.~Honda, T.~Kajita, K.~Kasahara, and S.~Midorikawa, {\it {Improvement of low
  energy atmospheric neutrino flux calculation using the JAM nuclear
  interaction model}},  {\em Phys.Rev.} {\bf D83} (2011) 123001,
  [\href{http://xxx.lanl.gov/abs/1102.2688}{{\tt arXiv:1102.2688}}].

\bibitem{Honda:2006qj}
M.~Honda, T.~Kajita, K.~Kasahara, S.~Midorikawa, and T.~Sanuki, {\it
  {Calculation of atmospheric neutrino flux using the interaction model
  calibrated with atmospheric muon data}},  {\em Phys.Rev.} {\bf D75} (2007)
  043006, [\href{http://xxx.lanl.gov/abs/astro-ph/0611418}{{\tt
  astro-ph/0611418}}].

\end{thebibliography}\endgroup

\end{document}